\documentclass[11pt,onecolumn,draft]{IEEEtran}
\usepackage{mathrsfs,pslatex}
\usepackage{indentfirst}
\usepackage{amssymb}
\usepackage{latexsym,amsfonts,amstext,makeidx,amsmath,amssymb,amsthm,color,threeparttable}
\newcommand{\Tr}{\text{Tr}}

\newtheorem{thm}{Theorem}
\newtheorem{prop}{Proposition}
\newtheorem{lem}{Lemma}
\newtheorem{cor}{Corollary}
\newtheorem{Rem}{Remark}

\begin{document}

\title{On the weight distributions of several classes of cyclic codes from APN monomials}
\author{Chunlei Li, Nian Li,
    Tor Helleseth 
    and Cunsheng Ding 
 \thanks{C. Li and T. Helleseth are with the Department of Informatics, University of Bergen, N-5020 Bergen, Norway
         (E-mail: chunlei.li@ii.uib.no, tor.helleseth@ii.uib.no).}
 \thanks{N. Li is with the Information Security and National Computing Grid Laboratory, Southwest Jiaotong
         University, Chengdu, 610031, China (E-mail: nianli.2010@gmail.com).}
  \thanks{C. Ding is with the Department of Computer Science and Engineering, The Hong Kong University of Science and Technology, Clear Water Bay, Kowloon, Hong Kong, China (E-mail: cding@ust.hk).}
}
\date{\today}
\maketitle

\begin{abstract}
Let $m\geq 3$ be an odd integer and $p$ be an odd prime. 
 In this paper,  many classes of three-weight cyclic codes over $\mathbb{F}_{p}$ are presented via
 an examination of the condition for the cyclic codes $\mathcal{C}_{(1,d)}$ and $\mathcal{C}_{(1,e)}$, which have parity-check polynomials $m_1(x)m_d(x)$ and $m_1(x)m_e(x)$ respectively, to have the same weight distribution,
where $m_i(x)$ is the minimal polynomial of $\pi^{-i}$ over $\mathbb{F}_{p}$ for a primitive element $\pi$ of $\mathbb{F}_{p^m}$.
Furthermore, for $p\equiv 3 \pmod{4}$ and positive integers $e$ such that there exist integers $k$ with $\gcd(m,k)=1$ and $\tau\in\{0,1,\cdots, m-1\}$ satisfying $(p^k+1)\cdot e\equiv 2 p^{\tau}\pmod{p^m-1}$,
the value distributions of the two exponential sums
$T(a,b)=\sum\limits_{x\in \mathbb{F}_{p^m}}\omega^{\Tr(ax+bx^e)}$
and
$
S(a,b,c)=\sum\limits_{x\in \mathbb{F}_{p^m}}\omega^{\Tr(ax+bx^e+cx^s)},
$
where $s=(p^m-1)/2$,  are settled.
As an application, the value distribution of $S(a,b,c)$ is utilized to investigate the weight distribution of the cyclic codes $\mathcal{C}_{(1,e,s)}$ with parity-check polynomial $m_1(x)m_e(x)m_s(x)$. In the case of $p=3$ and even $e$ satisfying the above condition, the duals of the
cyclic codes $\mathcal{C}_{(1,e,s)}$ have the optimal minimum distance.
\end{abstract}


\section{Introduction}\label{Sec1}

Let $p$ be a prime, $m$ be a positive integer and $q=p^m$. Let $\mathbb{F}_{q}$ denote the finite field with $q$ elements and $\mathbb{F}_{q}^*=\mathbb{F}_{q}\setminus\{0\}$. A linear $[n,\kappa, \rho]$ code $\mathcal{C}$ over $\mathbb{F}_{p}$ is a $\kappa$-dimensional subspace of $\mathbb{F}_{p}^n$ with minimum (Hamming)
nonzero weight $\rho$. Let $A_i$ denote the number of codewords in $\mathcal{C}$ with Hamming weight $i$.
The weight distribution $\{A_0,A_1,\cdots,A_n\}$ is an important research object in coding theory because it contains crucial information as to estimate the error correcting  capability and allows the computation
of the error probability of error detection and correction
with respect to some error detection and error correction
algorithms \cite{Klove}.

A linear code $\mathcal{C}$ over $\mathbb{F}_{p}$ is called {\em cyclic} if any cyclic shift of a codeword
is another codeword of $\mathcal{C}$.
 It is well known that any cyclic code of length $n$ over $\mathbb{F}_{p}$ corresponds to an ideal of the polynomial residue class ring $\mathbb{F}_{p}[x]/(x^n-1)$ and can be expressed as $\mathcal{C}=\langle g(x) \rangle$, where $g(x)$ is monic and has the least degree. This polynomial  $g(x)$ is called the {\em generator polynomial} and $h(x)=(x^n-1)/g(x)$ is referred to as the {\em parity-check polynomial} of $\mathcal{C}$. Cyclic codes with a few weights are of particular interest in
secret sharing schemes and designing frequency hopping sequences and
have been extensively studied in the literature (see, for example, \cite{Carlet-Ding-Yuan, Ding1, DingLiuMaZeng, Feng2008390, TFeng, LuoFeng2, Zeng-HJYC, Zeng-SH}). In this paper,
cyclic codes with $t$ nonzero weights are called $t$-weight cyclic codes.

Let $\Gamma_j$ be the $p$-cyclotomic coset modular $q-1$ containing $j$, i.e.,
$$\Gamma_j=\{j\cdot p^{i} \bmod{(q-1)} | i=0,1,\cdots, m-1\},$$
where $j$ is any integer with $0\leq j\leq q-2$. For
integers $i_1,\cdots,i_t\in\mathbb{Z}_{q-1}$, $t\geq 1$,  such that the cyclotomic cosets $\Gamma_{i_1},\cdots, \Gamma_{i_t}$ are pairwise disjoint,
 let $\mathcal{C}_{(i_1,\cdots, i_t)}$  denote the cyclic code with parity-check polynomial $h(x)=m_{i_1}(x)\cdots m_{i_t}(x)$ and write $\mathcal{C}_{(i_1,\cdots, i_t)}^{\bot}$ for its dual code, where and whereafter $\mathbb{Z}_{q-1}$ is the set of integers modulo $q-1$ and $m_i(x)$ is the minimal polynomial
 of $\pi^{-i}$ over $\mathbb{F}_p$.

 When $x^e$ is a perfect nonlinear (PN) monomial over $\mathbb{F}_{q}$,
 the properties of the cyclic codes
 $\mathcal{C}_{(1,e)}$ and $\mathcal{C}_{(0, 1,e)}$ and their dual codes were
studied in \cite{Carlet-Ding-Yuan, FengLuo, LiChao, YCD06}. It was shown that
 the cyclic codes
 $\mathcal{C}_{(1,e)}$ and $\mathcal{C}_{(0, 1,e)}$ have only a few weights and their dual codes have optimal minimum distances $4$ and $5$ respectively.
Very recently, for some monomials $x^e$ including all almost perfect nonlinear (APN) monomials,
the cyclic codes $\mathcal{C}_{(1,e)}^{\bot}$ were shown to have optimal minimum distance 4\cite{Ding-Helleseth}. Proceeding in this direction, the authors in \cite{LLHDT} further presented a number of monomials $x^e$ such that the corresponding cyclic codes $\mathcal{C}_{(1,e)}^{\bot}$ and
$\mathcal{C}_{(1,e,s)}^{\bot}$, where $s=(3^m-1)/2$, achieve the optimal minimum distance $4$ and $5$, respectively.


In this paper, for odd integer $m\geq 3$, we will derive general conditions on the parameters $e, p$ and $m$
under which $\mathcal{C}_{(1, e)}$ is a three-weight code.
It turns out that all the three-weight cyclic codes recently found in \cite{CKNC,Zhou1,Zhou2}
are special cases of the general construction of this paper and new three-weight cyclic codes are generated.
Furthermore,
for $p\equiv 3 \pmod{4}$ and positive integers $e$ such that there exist integers $k$ with $\gcd(m,k)=1$ and $\tau\in\mathbb{Z}_m$ satisfying $(p^k+1)\cdot e\equiv 2 p^{\tau}\pmod{p^m-1}$, we will determine
the value distributions of the two exponential sums
\begin{equation*}
T(a,b)=\sum\limits_{x\in \mathbb{F}_{p^m}}\omega^{\Tr(ax+bx^e)},\quad
S(a,b,c)=\sum\limits_{x\in \mathbb{F}_{p^m}}\omega^{\Tr(ax+bx^e+cx^s)},
\end{equation*} where $s=(p^m-1)/2$,  $\omega$ is the $p$-th root of unity and $\Tr(\cdot)$ is the trace mapping from $\mathbb{F}_{q}$ to $\mathbb{F}_p$.
The value distribution of $S(a,b,c)$ is subsequently utilized to
investigate the weight distribution of
the cyclic codes $\mathcal{C}_{(1,e,s)}$.
For $p=3$ and even $e$ satisfying $(p^k+1)\cdot e\equiv 2p^{\tau} \pmod{p^m-1}$ with $\gcd(2m,k)=1$,
the cyclic codes $\mathcal{C}_{(1,e,s)}^{\bot}$ were shown to have optimal minimum distance $5$ \cite{LLHDT}.

The remainder of this paper is organized as follows. Section \ref{Sec2} introduces some preliminary results. Section \ref{Sec2-1} presents
 a unified approach to generating three-weight cyclic codes, whose weight distributions are as well settled.
 Section \ref{Sec3} deals with the value distribution of the
 exponential sums $T(a,b)$ and $S(a,b,c)$. Section \ref{Sec4} determines the weight distribution of the cyclic code $\mathcal{C}_{(1,e,s)}$. Section \ref{sec-last} concludes this paper.

%

\section{Preliminaries}\label{Sec2}

A function $f: \mathbb{F}_{q} \to \mathbb{F}_{q}$ is referred to as
{\em perfect
nonlinear (PN)} if
$$
\max_{a \in \mathbb{F}_{q}^*} \max_{b \in \mathbb{F}_{q}} |\{x \in \mathbb{F}_{q}: f(x+a)-f(x)=b\}| =1
$$
and is called {\em almost perfect
nonlinear (APN)} if
$$
\max_{a \in \mathbb{F}_{q}^*} \max_{b \in \mathbb{F}_{q}} |\{x \in \mathbb{F}_{q}: f(x+a)-f(x)=b\}| =2.
$$
If a monomial $x^e$ is PN (APN) over $\mathbb{F}_{q}$, the exponent $e$ is referred to as {\em a PN (APN) exponent}.
PN and APN functions are two important research objects in cryptography and coding theory.
It is easily seen that there exists no PN function on $\mathbb{F}_{2^m}$. Much work has been done on
PN functions over $\mathbb{F}_{q}$ for odd prime $p$ \cite{Carlet-Ding, Coulter-Matthews}, and APN functions over $\mathbb{F}_{q}$ for arbitrary prime $p$
 \cite{Carlet, HRS, Zha}.


By the well-known Delsarte's Theorem \cite{Delsarte}, one can express the cyclic code $\mathcal{C}_{(i_1,\cdots,i_t)}$
as
$$
\mathcal{C}_{(i_1,\cdots,i_t)}=\left\{\Big(\sum_{s=1}^t\Tr(a_{s}\pi^{j\cdot i_s})\Big)_{j=0}^{q-2}\,\Big|\,a_1,a_2,\cdots, a_t\in \mathbb{F}_{q}\right\}.
$$Hence the Hamming weight of the codeword $\mathbf{c}=(c_0,c_1,\cdots,c_{q-2})$ in $\mathcal{C}_{(i_1,\cdots,i_t)}$ is
\begin{equation}\label{EqHammingWt}\begin{array}{rcl}
  w_H(\mathbf{c})&=&\big|\{j\,|\,0\leq j\leq q-2,\, c_j\neq 0\}\big|
  \\&=&(q-1)-\big|\{j\,|\,0\leq j\leq q-2,\, c_j=0\}\big|
  \\&=&(q-1)-\frac{1}{p}\sum\limits_{x\in \mathbb{F}_{q}^*}\sum\limits_{y\in \mathbb{F}_{p}}\omega^{y\cdot \Tr(\sum_{s=1}^ta_{s}x^{i_s})}
  \\&=&(q-1)-\frac{(q-1)}{p}-\frac{1}{p}\sum\limits_{y\in \mathbb{F}_{p}^*}\sum\limits_{x\in \mathbb{F}_{q}^*}\omega^{y\cdot \Tr(\sum_{s=1}^ta_{s}x^{i_s})}
  \\&=&\frac{(q-1)(p-1)}{p}-\frac{1}{p}\sum\limits_{y\in \mathbb{F}_{p}^*}\sum\limits_{x\in \mathbb{F}_{q}^*}\omega^{ \Tr(y\sum_{s=1}^ta_{s}x^{i_s})}
  \\&=&\frac{(q-1)(p-1)}{p}+\frac{(p-1)}{p}-\frac{1}{p}\sum\limits_{y\in \mathbb{F}_{p}^*}\sum\limits_{x\in \mathbb{F}_{q}}\omega^{ \Tr(y\sum_{s=1}^ta_{s}x^{i_s})}
  \\&=&p^{m-1}(p-1)-\frac{1}{p}\sum\limits_{y\in \mathbb{F}_{p}^*}S(ya_1,ya_2,\cdots,ya_l),
\end{array}\end{equation}where
$$
S(a_1,a_2,\cdots,a_t)=\sum\limits_{x\in \mathbb{F}_{q}}\omega^{\Tr(a_1x^{i_1}+a_2x^{i_2}+\cdots+a_lx^{i_t})}.
$$
In this way, the weight distribution of the cyclic code $\mathcal{C}_{(i_1,\cdots,i_t)}$ can be derived from determining
the following set:
$$
\{S(a_1,a_2,\cdots,a_t)\,|\, a_1,a_2,\cdots,a_t\in \mathbb{F}_{q}\}.
$$

In the subsequent sections, we will restrict ourselves to the weight distributions of cyclic codes $\mathcal{C}_{(1,e)}$ and $\mathcal{C}_{(1,e,s)}$, where $s=(p^m-1)/2$ for positive integers $e$ satisfying certain conditions.

\section{Three-weight cyclic codes and their weight distributions} \label{Sec2-1}

In this section our task is to derive general conditions on $(p, m, e)$ under which $\mathcal{C}_{(1,e)}$ is a three-weight code.
To this end, we need to introduce earlier results on three-weight cyclic codes $\mathcal{C}_{(1,e)}$.

In \cite{Carlet-Ding-Yuan, YCD06}, Carlet et al. employed PN monomials to construct three-weight cyclic codes
documented in the following lemma.

\begin{lem} \cite{Carlet-Ding-Yuan, YCD06}\label{Lem_Sec2-1}
   Let $p$ be an odd prime and $m\geq 3$ be odd. Then the cyclic code $\mathcal{C}_{(1,e)}$ has length $q-1$,
   dimension $2m$, and the weight distribution in Table \ref{Tab Sec2-1}
   if
\begin{enumerate}
  \item $e=p^k+1$ or
  \item $e=(p^k+1)/2$, where $p=3$ and $\gcd(2m,k)=1$.
\end{enumerate}
 \end{lem}

 The second construction in Lemma \ref{Lem_Sec2-1} was extended to any odd prime $p$  in \cite{LuoFeng2, Zhou}.

\begin{lem} \cite{LuoFeng2, Zhou}\label{Lem_Sec2-2}
 Let $p$ be an odd prime and $m\geq 3$ be odd. Let $k$ be a positive integer coprime to $m$ and $e=(p^k+1)/2$. Then the cyclic code $\mathcal{C}_{(1,e)}$ has length $q-1$, dimension $2m$, and the weight distribution in Table \ref{Tab Sec2-2} if $k$ is odd and  in Table \ref{Tab Sec2-3} if $k$ is even.
 \end{lem}

\begin{table}
\begin{center}
\caption{\small{Weight distribution I  }}
\label{Tab Sec2-1}
\begin{tabular}{|c|c|}
\hline Hamming weight & Multiplicity \\
\hline
$0$& $1$\\ \hline
$(p-1)p^{m-1}-p^{\frac{m-1}{2}}$ & $\frac{1}{2}(p-1)(p^m-1)(p^{m-1}+p^{\frac{m-1}{2}})$\\ \hline
$(p-1)p^{m-1}+p^{\frac{m-1}{2}}$ & $\frac{1}{2}(p-1)(p^m-1)(p^{m-1}-p^{\frac{m-1}{2}})$\\ \hline
$(p-1)p^{m-1}$ & $(p^m-1)(p^{m-1}+1)$\\
\hline
\end{tabular}
\end{center}
\end{table}

\begin{table}
\begin{center}
\caption{\small{Weight distribution II}}
\label{Tab Sec2-2}
\begin{tabular}{|c|c|}
\hline Hamming weight & Multiplicity \\
\hline
$0$& $1$\\ \hline
$(p-1)p^{m-1}-\frac{(p-1)}{2}p^{\frac{m-1}{2}}$ & $(p^m-1)(p^{m-1}+p^{\frac{m-1}{2}})$\\ \hline
$(p-1)p^{m-1}+\frac{(p-1)}{2}p^{\frac{m-1}{2}}$ & $(p^m-1)(p^{m-1}-p^{\frac{m-1}{2}})$\\ \hline
$(p-1)p^{m-1}$ & $(p^m-1)(p^m-2p^{m-1}+1)$\\
\hline
\end{tabular}
\end{center}
\end{table}

\begin{table}
\begin{center}
\caption{\small{Weight distribution III}}
\label{Tab Sec2-3}
\begin{tabular}{|c|c|}
\hline Hamming weight & Multiplicity \\
\hline
$0$& $1$\\ \hline
$(p-1)p^{m-1}-(p-1)p^{\frac{m-1}{2}}$ & $\frac{1}{2}(p^m-1)(p^{m-1}+p^{\frac{m-1}{2}})$\\ \hline
$(p-1)p^{m-1}+(p-1)p^{\frac{m-1}{2}}$ & $\frac{1}{2}(p^m-1)(p^{m-1}-p^{\frac{m-1}{2}})$\\ \hline
$(p-1)p^{m-1}$ & $(p^m-1)(p^m-p^{m-1}+1)$\\
\hline
\end{tabular}
\end{center}
\end{table}

 The following lemma will be needed in the sequel.

 \begin{lem}  \label{Lem_Sec2-4}
   Let $m\geq 3$ be odd and $p$ be an odd prime with $p-1=2^r\cdot h$, where $h$ is an odd integer.
   If two integers $d, \, e\in\mathbb{Z}_{p^m-1}\setminus\Gamma_{1}$ satisfy $2de\equiv 2p^{\tau} \pmod{p^m-1}$ for some $\tau\in\mathbb{Z}_m$ and $d+e\equiv 2 \pmod{2^r}$, then
   the cyclic codes $\mathcal{C}_{(1,d)}$ and $\mathcal{C}_{(1, e)}$ have the same weight distribution.
 \end{lem}

\noindent\textbf{Proof.}
Let $e_1\equiv e\cdot p^{m-\tau}  \pmod{p^m-1}$. Then the integers $d$ and $e_1$ satisfy $2de_1\equiv 2 \pmod{p^m-1}$ and $d+e_1\equiv 2 \pmod{2^r}$.
Thus, it is sufficient to show that the assertion holds for the case $\tau=0$ since the cyclic codes $\mathcal{C}_{(1,e)}$ and
$\mathcal{C}_{(1,e_1)}$ are the same.

According to (\ref{EqHammingWt}),  the weight distributions of $\mathcal{C}_{(1,d)}$ and $\mathcal{C}_{(1, e)}$
are respectively determined by the value distributions of
$$
\Delta_0(a,b)=\sum\limits_{y\in \mathbb{F}_{p}^*}\sum\limits_{x\in \mathbb{F}_{q}}\omega^{\Tr(yax+ybx^d)}
\text{ and }
\Delta_1(a,b)=\sum\limits_{y\in \mathbb{F}_{p}^*}\sum\limits_{x\in \mathbb{F}_{q}}\omega^{\Tr(yax+ybx^e)}.
$$
Notice that $h$ and $m$ are odd.
The element $\lambda=\pi^{\frac{(p^m-1)h}{p-1}}$, where $\pi$ is a primitive element in $\mathbb{F}_{p^m}$, is
a non-square in $\mathbb{F}_{p^m}$. It then follows from $p-1=2^r\cdot h$ that the order of $\lambda$ in $\mathbb{F}_{p}^*$ (the least integer $t$ such that $\lambda^t=1$) equals $2^r$ .

When $x$ runs through $\mathbb{F}_{q}^*$, $x^{2}$ runs twice through the
squares in $\mathbb{F}_{q}^*$, and $\lambda x^{2}$ runs twice through all the
non-squares in $\mathbb{F}_{q}^*$. Thus,
$$
\begin{array}{rcl}
\Delta_0(a,b)&=&\frac{1}{2}\sum\limits_{y\in \mathbb{F}_{p}^*}\big(\sum\limits_{x\in \mathbb{F}_{q}}\omega^{\Tr(yax^{2}+ybx^{2d})}+\sum\limits_{x\in \mathbb{F}_{q}}\omega^{\Tr(ya\lambda x^{2}+yb\lambda^d x^{2d})}\big)
\\&=&
\frac{1}{2}\Big(\sum\limits_{y\in \mathbb{F}_{p}^*}\sum\limits_{x\in \mathbb{F}_{q}}\omega^{y\Tr(ax^2+bx^{2d})}+\sum\limits_{y\in \mathbb{F}_{p}^*}\sum\limits_{x\in \mathbb{F}_{q}}\omega^{y\lambda\Tr(ax^{2}+\lambda^{d-1}bx^{2d})}\Big)
\\&=&
\frac{1}{2}\Big(\sum\limits_{y\in \mathbb{F}_{p}^*}\sum\limits_{x\in \mathbb{F}_{q}}\omega^{y\Tr(ax^2+bx^{2d})}+\sum\limits_{y\in \mathbb{F}_{p}^*}\sum\limits_{x\in \mathbb{F}_{q}}\omega^{y\Tr(ax^{2}+\lambda^{d-1}bx^{2d})}\Big).
\end{array}
$$

Note that $2de\equiv 2 \pmod{p^m-1}$ implies $\gcd(2d,p^m-1)=2$.  Thus, when $x$ runs through $\mathbb{F}_{q}^*$, $x^{2d}$ runs twice through the
squares in $\mathbb{F}_{q}^*$, and $\lambda x^{2d}$ runs twice through all the
non-squares in $\mathbb{F}_{q}^*$.
Similarly we have
$$\begin{array}{rcl}
\Delta_1(a,b)&=&
\frac{1}{2}\Big(\sum\limits_{y\in \mathbb{F}_{p}^*}\sum\limits_{x\in \mathbb{F}_{q}}\omega^{y\Tr(ax^{2d}+bx^2)}+\sum\limits_{y\in \mathbb{F}_{p}^*}\sum\limits_{x\in \mathbb{F}_{q}}\omega^{y\Tr(ax^{2d}+\lambda^{e-1}bx^2)}\Big).
\end{array}
$$
Furthermore, it follows from $d+e\equiv 2  \pmod{2^r}$ and $\lambda^{2^r}=1$ that
$$
\begin{array}{rcl}
\Delta_0(a,b)&=&\frac{1}{2}\Big(\sum\limits_{y\in \mathbb{F}_{p}^*}\sum\limits_{x\in \mathbb{F}_{q}}\omega^{y\Tr(ax^2+bx^{2d})}+\sum\limits_{y\in \mathbb{F}_{p}^*}\sum\limits_{x\in \mathbb{F}_{q}}\omega^{y\Tr(\lambda^{e+d-2}ax^{2}+\lambda^{d-1}bx^{2d})}\Big)
\\&=&\frac{1}{2}\Big(\sum\limits_{y\in \mathbb{F}_{p}^*}\sum\limits_{x\in \mathbb{F}_{q}}\omega^{y\Tr(ax^2+bx^{2d})}+\sum\limits_{y\in \mathbb{F}_{p}^*}\sum\limits_{x\in \mathbb{F}_{q}}\omega^{y\lambda^{d-1}\Tr(\lambda^{e-1}ax^{2}+bx^{2d})}\Big)
\\&=&\frac{1}{2}\Big(\sum\limits_{y\in \mathbb{F}_{p}^*}\sum\limits_{x\in \mathbb{F}_{q}}\omega^{y\Tr(ax^2+bx^{2d})}+\sum\limits_{y\in \mathbb{F}_{p}^*}\sum\limits_{x\in \mathbb{F}_{q}}\omega^{y\Tr(\lambda^{e-1}ax^{2}+bx^{2d})}\Big)
\\&=&\Delta_1(b,a).
\end{array}
$$
Therefore, the multi-sets  $\{\Delta_0(a,b):\, a, b\in \mathbb{F}_{q}\}$ and $\{\Delta_1(a,b):\, a, b\in \mathbb{F}_{q}\}$ have the
same value distribution.
\hfill$\blacksquare$



\smallskip

Applying Lemmas \ref{Lem_Sec2-1}-\ref{Lem_Sec2-4}, we obtain the following.

\begin{thm}\label{Thm1}
Let $m \geq 3$ be odd.
(i) Let $p \equiv 3 \pmod{4}$. If $e$ is an even integer satisfying $2(p^k+1)e\equiv 2p^{\tau}  \pmod{p^m-1}$  for some
$\tau \in \mathbb{Z}_m$ and some positive integer $k$, then $\mathcal{C}_{(1,e)}$ is a three-weight cyclic code with the weight distribution in Table \ref{Tab Sec2-1}.
(ii) Let $p$ be any odd prime. If $e$ is an integer satisfying $(p^k+1)e\equiv 2p^{\tau} \pmod{p^m-1}$ for some
$\tau \in \mathbb{Z}_m$ and some positive integer $k$ with $\gcd(m,k)=1$, then $\mathcal{C}_{(1,e)}$ is a three-weight
cyclic code with the weight distribution of
\begin{itemize}
\item Table \ref{Tab Sec2-2} when $e \equiv 1 + (p-1)/2 \pmod{p-1}$; and
\item Table \ref{Tab Sec2-3} when  $e \equiv 1 \pmod{p-1}$.
\end{itemize}

\end{thm}

\noindent\textbf{Proof.}
(i) When $m$ is odd and $p \equiv 3 \pmod{4}$, we have $p-1=2^r h$ with $r=1$ and $h=(p-1)/2$ being odd.
Set $d=p^k+1$. By assumption, $2de \equiv 2p^{\tau}  \pmod{p^m-1}$ for some $\tau$. In addition, we have
$e+d \equiv 2 \pmod{2^r}$ as $e$ is even by assumption and $d$ is obviously even.

Since $m$ is odd and $p \equiv 3 \pmod{4}$,  $p^m\equiv 3 \pmod{4}$. Hence  $\gcd(2(p^k+1),p^m-1)=2$.
It then follows that $e \not\in \Gamma_1$. It is clear that $d \not\in \Gamma_1$. Thus, all the conditions in
Lemma \ref{Lem_Sec2-4} are satisfied.
Then it follows from Lemmas \ref{Lem_Sec2-1} and \ref{Lem_Sec2-4} that
$\mathcal{C}_{(1,e)}$ has the weight distribution of Table \ref{Tab Sec2-1}.

(ii) Let $p$ be any odd prime and let $m$ be odd.  Assume that $e$ is an integer satisfying $(p^k+1)e\equiv 2p^{\tau}
\pmod{p^m-1}$ for some $\tau\in\mathbb{Z}_m$ and some positive integer $k$ with $\gcd(m,k)=1$. One can then
easily prove that $\gcd(p^k+1,p^m-1)=2$ and then $e \not\in \Gamma_1$.

By assumption, $(p^k+1)e \equiv 2p^{\tau} \pmod{p^m-1}$, which implies that $2e \equiv 2 \pmod{p-1}$.
Thus one has either $e\equiv 1 \pmod{p-1}$ or $e \equiv 1+(p-1)/2 \pmod{p-1}$.

Note that $(p^k+1)/2\equiv 1+(p^k-1)/2\equiv 1+k(p-1)/2 \pmod{p-1}$. In the case that $e\equiv 1 \pmod{p-1}$
 and $k$ is odd, one has
 $e+(p^{m-k}+1)/2\equiv 2 \pmod{p-1}$.  Since $(p^{m-k}+1)e\equiv 2p^{\tau+m-k} \pmod{p^m-1}$,
  it follows from Lemma \ref{Lem_Sec2-4} that
$\mathcal{C}_{(1,e)}$ has the same weight distribution as $\mathcal{C}_{(1,d)}$ for $d=(p^{m-k}+1)/2$. In the case that
 $e\equiv 1  \pmod{p-1}$ and $k$ is even, one has $e+(p^{k}+1)/2\equiv 2  \pmod{p-1}$ and then
$\mathcal{C}_{(1,e)}$ has the same weight distribution as $\mathcal{C}_{(1,d)}$ for $d=(p^{k}+1)/2$.
Therefore,  when $e\equiv 1 \pmod{p-1}$,  it follows from Lemma \ref{Lem_Sec2-2} that $\mathcal{C}_{(1,e)}$ has
the weight distribution of Table \ref{Tab Sec2-3}. Similarly, one can prove that $\mathcal{C}_{(1,e)}$ has
the weight distribution of Table \ref{Tab Sec2-2} when $e \equiv 1+(p-1)/2 \pmod{p-1}$. The proof is completed. \hfill$\blacksquare$

\smallskip

\begin{Rem}\label{Rem1}
{\rm
Very recently, a total of thirteen classes of three-weight $[p^m-1, 2m]$ cyclic codes over $\mathbb{F}_{p}$ are
described in \cite{CKNC,Zhou1,Zhou2}. It can be verified by hand
that all the three-weight cyclic codes found in \cite{CKNC,Zhou1, Zhou2} are special cases of the codes in Theorem \ref{Thm1}.
Furthermore, a closer look at Theorem \ref{Thm1} shows that
one can derive in total $2\phi(m)$ three-weight cyclic codes $\mathcal{C}_{(1,e)}$ for $d=(p^k+1)/2$, $\gcd(k,m)=1$ and $p\equiv 1 \pmod{4}$, and
$m+2\phi(m)$ three-weight cyclic codes $\mathcal{C}_{(1,e)}$ from $d \in \{p^k+1, (p^k+1)/2\}$, $\gcd(k,m)=1$ and $p\equiv 3 \pmod{4}$, where $\phi(\cdot)$ is
the Euler's phi function.
}
\end{Rem}

We end this section by listing some special integers $e$ in Theorem \ref{Thm1}, which are APN exponents and
were employed to
generate optimal cyclic code $\mathcal{C}^{\bot}_{(1,e)}$ and $\mathcal{C}^{\bot}_{(1,e,s)}$ in \cite{Ding-Helleseth, LLHDT}.
\begin{enumerate}
      \item $e=\frac{3^{(m+1)/2}-1}{2}$ for $m\equiv 3 \pmod{4}$; and $e=\frac{3^{(m+1)/2}-1}{2}+\frac{3^m-1}{2}$ for $m\equiv 1 \pmod{4}$;
      \item $e=\frac{3^{m+1}-1}{8}$ for $m\equiv 3 \pmod{4}$; and $e=\frac{3^{m+1}-1}{8}+\frac{3^m-1}{2}$ for $m\equiv 1 \pmod{4}$;
      \item $e=\frac{3^{m}+1}{4}+\frac{3^m-1}{2}$;
      \item $e=3^{(m+1)/2}-1$;
      \item $e=(3^{(m+1)/4}-1)(3^{(m+1)/2}+1)$ for $m\equiv 3 \pmod{4}$,
\end{enumerate} where the corresponding integers $d$ are:
\begin{enumerate}
      \item $d=3^k+1$ with $k=(m+1)/2$;
      \item $d=3^k+1$ with $k=1$;
      \item $d=(3^k+1)/2$ with $k=1$;
      \item $d=(3^k+1)/2$ with $k=(m+1)/2$ if $m\equiv 1 \pmod{4}$ and $k=(m-1)/2$ if $m\equiv 3 \pmod{4}$;
      \item $d=(3^k+1)/2$ with $k=(m+1)/4$ if $m\equiv 3 \pmod{8}$ and $k=(3m-1)/4$ if $m\equiv 7 \pmod{8}$.
\end{enumerate}
For the APN exponents listed above, Theorem \ref{Thm1} shows that the cyclic codes $\mathcal{C}_{(1,e)}$ have the
weight distribution of Table \ref{Tab Sec2-1} (or Table \ref{Tab Sec2-2}) for $p=3$.


\section{Value distributions of the two exponential sums} \label{Sec3}

For odd $m \geq 3$, the weight distribution of the cyclic code $\mathcal{C}_{(1,e)}$ is determined in Theorem \ref{Thm1}
when the integer $e$ satisfies
$(p^k+1)e\equiv 2p^{\tau} \pmod{p^m-1}$ for some $\tau\in \mathbb{Z}_m$ and some positive integer $k$ with $\gcd(m,k)=1$.
In this section, for $p\equiv 3 \pmod{4}$,
we further study the following multi-sets
\begin{equation}\label{EqExpSumT}
\Big\{T(a,b)=\sum\limits_{x\in \mathbb{F}_{q}}\omega^{\Tr(ax+bx^e)}: a,\,b\in \mathbb{F}_{q}\Big\}
\end{equation}
and
\begin{equation}\label{EqExpSumS}
\Big\{S(a,b,c)=\sum\limits_{x\in \mathbb{F}_{q}}\omega^{\Tr(ax+bx^e+cx^s)}: a, \,b \in \mathbb{F}_{q}, c\in \Omega \Big\},
\end{equation} where $s=(p^m-1)/2$ and $\Omega=\{\omega_0,\omega_1,\cdots,\omega_{p-1}\}$ such that $\{\Tr(\omega)|\omega\in \Omega\}=\mathbb{F}_{p}$.
The value distribution of $S(a,b,c)$
will be utilized to
investigate the weight distribution of
the cyclic codes $\mathcal{C}_{(1,e,s)}$ in Section \ref{Sec4}.

For convenience and ease of presentation in the sequel, an integer $e$ is hereafter said to satisfy the {\em Congruence Condition} if there exist integers $k$ with $\gcd(m,k)=1$, $\tau\in \mathbb{Z}_m$ such that $(p^k+1)e \equiv 2p^{\tau} \pmod{p^m-1}$.
It is worth noting that for a positive integer $e$, 
the multisets $\{T(a,b):\, a, b\in \mathbb{F}_{q}\}$ and $\{S(a,b,c): a, \,b \in \mathbb{F}_{q}, c\in \Omega \}$
remain unchanged when $e$ runs through the integers in $\Gamma_e$. This fact allows us to assume $\tau=0$ throughout what follows.

For odd $m$ and $p\equiv 3 \pmod{4}$, $-1$ is a non-square in $\mathbb{F}_{q}$. Thus,
when $x$ runs through $\mathbb{F}_{q}^*$, $x^{p^k+1}$ runs twice through the
squares in $\mathbb{F}_{q}^*$ and $-x^{p^k+1}$ runs twice through all the
non-squares in $\mathbb{F}_{q}^*$.
Therefore, for integers $e$ satisfying the Congruence Condition, the exponential sums $T(a,b)$
and $S(a,b,c)$ can be rewritten as
\begin{equation}\label{EqExpSum3}\begin{array}{c}
T(a,b)=\frac{1}{2}\Big(\sum\limits_{x\in \mathbb{F}_{q}}\omega^{\Tr(ax^{p^k+1}+bx^{2})}+\sum\limits_{x\in \mathbb{F}_{q}}\omega^{\Tr(-ax^{p^k+1}+(-1)^ebx^{2})}\Big)
\end{array}\end{equation}
and
\begin{equation}\label{EqExpSum4}\begin{array}{c}
S(a,b,c)=1+\frac{1}{2}\Big(\sum\limits_{x\in \mathbb{F}_{q}^*}\omega^{\Tr(ax^{p^k+1}+bx^2+c)}+\sum\limits_{x\in \mathbb{F}_{q}^*}\omega^{\Tr(-ax^{p^k+1}+(-1)^ebx^2-c)}\Big).
\end{array}\end{equation}
Let
\begin{equation}\label{EqT0}
    T_0(a,b)=\sum\limits_{x\in \mathbb{F}_{q}}\omega^{\Tr(ax^{p^k+1}+bx^2)},
\end{equation}
then
\begin{equation}\label{EqT}
    T(a,b)=\frac{1}{2}\big(T_0(a,b)+T_0(-a,(-1)^eb)\big),
    \end{equation}
and
\begin{equation}\label{EqS}
    S(a,b,c)=1-\frac{1}{2}\big(\omega^{t}+\omega^{-t}\big)+\frac{1}{2}\big(\omega^{t}T_0(a,b)+\omega^{-t}T_0(-a,(-1)^eb)\big),
\end{equation} where $t=\Tr(c)$.

\begin{table}
\begin{center}
\caption{\small{The value distribution of $T_0(a,b)$ for odd $m\geq 3$}}\label{Tab1}
\begin{tabular}{|c|c|}
\hline Values & Multiplicity (each)\\
\hline
$\pm \sqrt{p^*}p^{\frac{m-1}{2}}$ & $p^2(p^m-1)(p^m-p^{m-1}-p^{m-2}+1)/(2(p^2-1))$\\ \hline
$\pm p^{\frac{m+1}{2}}$ & $(p^m-1)(p^{m-1}\pm p^{\frac{m-1}{2}})/2$\\ \hline
$\pm \sqrt{p^*}p^{\frac{m+1}{2}}$ & $(p^m-1)(p^{m-1}-1)/(2(p^2-1))$\\ \hline
$p^m$ & 1 \\ \hline
\end{tabular} 
\end{center}
\end{table}

Feng and Luo determined the value distribution of the exponential sum $T_0(a,b)$ when $(a,b)$ goes
through $\mathbb{F}_{q}^2$ \cite{Feng2008390}. The value distribution for odd $m\geq 3$
is given in Table \ref{Tab1}.
In order to determine the value distribution of $T(a,b)$ and $S(a,b,c)$, we
shall study the distribution of $\big(T_0(a,b),T_0(-a,(-1)^eb)\big)$ when  $(a,b)$ runs through
$\mathbb{F}_{q}^2$.

When $e$ is an odd integer, as $T_0(-a,-b)$ is the conjugate of $T_0(a,b)$ for any $(a,b)\in \mathbb{F}_{q}^2$,
the distribution of $\big(T_0(a,b),T_0(-a,-b)\big)$ can be readily settled from Table \ref{Tab1}.
We next focus our attention on the distribution of $\big(T_0(a,b),T_0(-a,b)\big)$.

The following two lemmas characterize all possible $\big(T_0(a,b),T_0(-a,b)\big)$ for any $a,b\in \mathbb{F}_{q}$.

\begin{lem}\label{Lem1}\cite{Feng2008390}
Let $Q(x)$ be a quadratic form in $m$ variables over $\mathbb{F}_{p}$ of rank $r$, $\big(\frac{a}{p}\big)$ be the conventional Legendre symbol. Then
$$
\sum_{x\in \mathbb{F}_{q}} \omega^{Q(x)}=\left\{
\begin{array}{cl}
\big(\frac{\Delta}{p}\big)p^{m-r/2}, & \text{if } p\equiv 1 \pmod{4} \\
(-1)^{r/2}\big(\frac{\Delta}{p}\big)p^{m-r/2}, & \text{if } p\equiv 3 \pmod{4},
\end{array}
\right.
$$ where $\Delta$ is the determinant of $Q(x)$. Furthermore, for any $y\in \mathbb{F}_{p}^*$,
\begin{equation}\label{EqLem1}
  \sum_{x\in \mathbb{F}_{q}} \omega^{yQ(x)}= \big(\frac{y^r}{p}\big)\sum_{x\in \mathbb{F}_{q}} \omega^{Q(x)}.
\end{equation}
\end{lem}


\begin{lem}\label{Lem1-2}\cite{Feng2008390} \cite{Zhou}
Let $m$ and $k$ be positive integers such that $\gcd(m,k)=1$. Let
$$Q_{a,b}(x)=\Tr(ax^{p^k+1}+bx^2)$$
be a quadratic form  in $m$ variables over $\mathbb{F}_{p}$. Then,
(i) for $(a,b)\in \mathbb{F}_{p^m}^2\setminus\{(0,0)\}$, the quadratic form $Q_{a,b}(x)$ has rank no less than $m-2$;
(ii) if $m$ is odd,
then for any $a\in \mathbb{F}_{p^m}^*$ and $b\in \mathbb{F}_{p^m}$, at least one of $Q_{a,b}(x)$ and $Q_{-a,b}(x)$ has rank $m$.
\end{lem}

For $i=0,1,2$, let
\begin{equation}\label{EqValue}
    \nu_i=\left\{
            \begin{array}{ll}
              p^{\frac{m+i}{2}}, & \hbox{if $i$ is odd,} \\
              \sqrt{p^*}p^{\frac{m+i}{2}}, & \hbox{if $i$ is even},
            \end{array}
          \right.
\end{equation}where $p^*=(-1)^{\frac{p-1}{2}}p$. By Lemmas \ref{Lem1} and \ref{Lem1-2}, for any $(a,b)\in \mathbb{F}_{q}^2\setminus\{(0,0)\}$,
$$\big(T_0(a,b),T_0(-a,b)\big)\in\big\{(\varepsilon_1\nu_{i_1},\varepsilon_2\nu_{i_2})\,|\, 0\leq i_1, i_2\leq 2, \varepsilon_1, \varepsilon_2=\pm 1\big\}.$$
Before further studying the value distribution of
$\big(T_0(a,b),T_0(-a,b)\big)$, we
define
\begin{equation}\label{EqSets}
\begin{array}{l}
N^{+}_{\varepsilon,i}=\Big\{(a,b)\in \mathbb{F}_{q}^2\,|\,T_0(a,b)=\varepsilon\nu_i\Big\}, \\
N^{-}_{\varepsilon,i}=\Big\{(a,b)\in \mathbb{F}_{q}^2\,|\,T_0(-a,b)=\varepsilon\nu_i\Big\},
\end{array}
\end{equation}  where $\varepsilon\in\{1,-1\}$.
Some properties of
$N^{+}_{\varepsilon,i}$ and $N^{-}_{\varepsilon,i}$ are summarized in the following lemma.
\begin{lem}\label{Lem3}
  Let $\lambda$ be a non-square of $\mathbb{F}_{p}^*$. For $\varepsilon\in\{1,-1\}$ and $i\in\{0,1,2\}$, we have
\begin{enumerate}
  \item[(i)]\label{P1} $(a,b)\in N^{+}_{\varepsilon,i}$ if and only if $(-a,b)\in N^{-}_{\varepsilon,i}$;
  \item[(ii)]\label{P2} $\lambda N^{+}_{\varepsilon,0}=N^{+}_{-\varepsilon,0}$, $\lambda N^{-}_{\varepsilon,0}=N^{-}_{-\varepsilon,0}$;
  \item[(iii)]\label{P3} $\lambda N^{+}_{\varepsilon,1}=N^{+}_{\varepsilon,1}$, $\lambda N^{-}_{\varepsilon,1}=N^{-}_{\varepsilon,1}$; and
  \item[(iv)]\label{P4} $\lambda N^{+}_{\varepsilon,2}=N^{+}_{-\varepsilon,2}$, $\lambda N^{-}_{\varepsilon,2}=N^{-}_{-\varepsilon,2}$.
\end{enumerate}
\end{lem}

\noindent\textbf{Proof.}
Property (i) directly follows from the definitions of
$N^{+}_{\varepsilon,i}$ and $N^{-}_{\varepsilon,i}$ in
(\ref{EqSets}). Properties (ii), (iii) and (iv) are proved
together below.

By Lemma \ref{Lem1}, for any non-square $\lambda$ of
$\mathbb{F}_{p}^*$,
\begin{equation}\label{EqT0lambda}
T_0(\lambda a, \lambda b)
= \sum_{x\in\mathbb{F}_{q}}\omega^{\lambda \Tr(
ax^{p^k+1}+b x^2)} =\Big(\frac{\lambda^{r}}{p}\Big) T_0(a,b),
\end{equation} where $r$ is the rank of the quadratic form
$Q_{a,b}(x)=\Tr(ax^{p^k+1}+bx^2)$. Following from the definitions of
$\nu_i$ and $N^{+}_{\varepsilon,i}$, we know that if $(a,b)\in
N^{+}_{\varepsilon,i}$, then the corresponding quadratic form
$Q_{a,b}(x)$ has rank $m-i$. This fact together with
(\ref{EqT0lambda}) implies that $T_0(\lambda a, \lambda b)=T_0(a,
b)$ if $(a,b)\in N^{+}_{\varepsilon,1}$ and $T_0(\lambda a, \lambda
b)=-T_0(a, b)$ if $(a,b)\in N^{+}_{\varepsilon,0}$ or $(a,b)\in
N^{+}_{\varepsilon,2}$. Therefore, we deduce that
$$
\lambda N^{+}_{\varepsilon,0}=N^{+}_{-\varepsilon,0}, \ \
 \lambda N^{+}_{\varepsilon,1}=N^{+}_{\varepsilon,1}, \ \
 \lambda N^{+}_{\varepsilon,2}=N^{+}_{-\varepsilon,2}.
 $$
 Then the properties for
$N^{-}_{\varepsilon, i}$  in (ii), (iii) and (iv) directly follow
from (i). \hfill$\blacksquare$

\smallskip

The following results are necessary for calculating the distribution of $\big(T_0(a,b),T_0(-a,b)\big)$.

\begin{prop}\label{prop2-1}
Let $\mathcal{N}_4$ denote the number of tuples $(x,y,z,w)\in \mathbb{F}_{q}^4$ satisfying
$$
\left\{\begin{array}{l}
x^{2}+y^{2}+z^{2}+w^{2}=0
\\
x^{p^k+1}+y^{p^k+1}+z^{p^k+1}-w^{p^k+1}=0.
\end{array}\right.
$$ Then for odd $m\geq 3$ and positive integer $k$ with $\gcd(m,k)=1$,
$$
\mathcal{N}_4=2q^2-qp-q+p.
$$
\end{prop}

\noindent\textbf{Proof.} See the Appendix for the details.

\begin{prop}\label{prop2} For odd $m\geq 3$ and positive integer $k$ with $\gcd(m,k)=1$,
\begin{enumerate}
  \item[(i)] $\sum\limits_{a,b\in \mathbb{F}_{q}}T_0(a,b)\cdot
  T_0(-a,b)=q^{2}$;
  \item[(ii)] $\sum\limits_{a,b\in \mathbb{F}_{q}}T^3_0(a,b)\cdot T_0(-a,b)=q^{2}\cdot(2q^2-qp-q+p)$.
\end{enumerate}
\end{prop}

\noindent\textbf{Proof.}
(i) By (\ref{EqT0}), one has
$$\begin{array}{rcl}
\lefteqn{\sum\limits_{a,\,b\in \mathbb{F}_{q}}T_0(a,b)\cdot T_0(-a,b)} \\
&=& \sum\limits_{a,\,b\in \mathbb{F}_{q}}\sum\limits_{x,\,y\in
\mathbb{F}_{q}}
\omega^{\Tr\big(a(x^{p^k+1}-y^{p^k+1})+b(x^2+y^2)\big)}
\\&=& \sum\limits_{x,\,y\in \mathbb{F}_{q}}\sum\limits_{a\in \mathbb{F}_{q}}
\omega^{\Tr(a(x^{p^k+1}-y^{p^k+1}))}\sum\limits_{b\in
\mathbb{F}_{q}} \omega^{\Tr(b(x^{2}+y^{2}))}
\\&=& q^2 \mathcal{N}_2,
\end{array}$$ where $\mathcal{N}_2$ is the number of the solutions
of the following system of equations:
$$
\left\{\begin{array}{l}
x^2+y^2=0\\
x^{p^k+1}-y^{p^k+1}=0.
\end{array}\right.
$$
Since $x^{p^k+1}=y^{p^k+1}$ and $\gcd(p^m-1,p^k+1)=2$, one has
$x^2=y^2$. This together with $x^2+y^2=0$ implies $x=y=0$, i.e.,
$\mathcal{N}_2=1$.

(ii)
In a similar manner, we deduce that
$$\begin{array}{rcl}
\lefteqn{\sum\limits_{a,\,b\in \mathbb{F}_{q}}T_0^3(a,b)\cdot T_0(-a,b)} \\
&=& \sum\limits_{x,\,y,z,\,w\in \mathbb{F}_{q}}\sum\limits_{a\in \mathbb{F}_{q}}
\omega^{\Tr(a(x^{p^k+1}+y^{p^k+1}+z^{p^k+1}-w^{p^k+1}))}\sum\limits_{b\in
\mathbb{F}_{q}} \omega^{\Tr(b(x^{2}+y^{2}+z^{2}+w^{2}))}
\\&=& p^{2m} \mathcal{N}_4,
\end{array}$$ where $\mathcal{N}_4$ is the number of the solutions
of the following system of equations:
$$
\left\{\begin{array}{l}
x^{2}+y^{2}+z^{2}+w^{2}=0
\\
x^{p^k+1}+y^{p^k+1}+z^{p^k+1}-w^{p^k+1}=0.
\end{array}\right.
$$ The proof immediately follows from Proposition \ref{prop2-1}.  \hfill$\blacksquare$

\medskip

With the preparations of Table \ref{Tab1}, Lemmas \ref{Lem1-2}, \ref{Lem3} and
Proposition \ref{prop2}, we are now ready to determine the
distribution of $\big(T_0(a,b),T_0(-a,b)\big)$.

\begin{thm}\label{ThmCoreResult} Let $v_i$, $i=0,1,2$, be defined by (\ref{EqValue}).
For odd $m\geq 3$ and positive integer $k$ with $\gcd(m,k)=1$, the distribution of the
multi-set
$$\Big\{\big(T_0(a,b),T_0(-a,b)\big)|a,b\in \mathbb{F}_{q}\Big\}$$ is shown
  in Table \ref{Tab2}.
\end{thm}

\begin{table}
\begin{center}
\caption{\small{The value distribution of $\big(T_0(a,b),T_0(-a,b)\big)$ for odd $m\geq 3$}}\label{Tab2}
\begin{tabular}{|c|c|}
\hline Values & Multiplicity (each)\\
\hline
$(\nu_0,\nu_0)$, $(-\nu_0,-\nu_0)$ & $(p^m-1)(p^{m+1}-3p^m+p+1)/(4(p-1))$ \\ \hline
$(\nu_0, -\nu_0)$, $(-\nu_0,\nu_0)$& $(p-1)(p^{2m}-1)/(4(p+1))$ \\ \hline
$(\nu_0,\nu_1)$, $(\nu_1,\nu_0)$ & $(p^m-1)(p^{m-1}+p^{\frac{m-1}{2}})/4$\\
$(-\nu_0,\nu_1)$, $(\nu_1,-\nu_0)$ &\\ \hline
$(\nu_0,-\nu_1)$, $(-\nu_1,\nu_0)$ & $(p^m-1)(p^{m-1}-p^{\frac{m-1}{2}})/4$\\
$(-\nu_0,-\nu_1)$, $(-\nu_1,-\nu_0)$ & \\ \hline
$(\nu_0,\nu_2)$, $(\nu_2,\nu_0)$ & ${(p^m-1)(p^{m-1}-1) }/(2 (p^2-1))$\\
$(-\nu_0,-\nu_2)$, $(-\nu_2,-\nu_0)$&  \\ \hline
$(\nu_0,-\nu_2)$, $(-\nu_2,\nu_0)$& $0$\\
$(-\nu_0,\nu_2)$, $(\nu_2,-\nu_0)$ & \\ \hline
$(p^m,p^m)$ & 1  \\
\hline
\end{tabular}
\end{center}
\end{table}

\noindent\textbf{Proof.}
By the definitions of $N^+_{\varepsilon,i}$ and $N^-_{\varepsilon,i}$ in (\ref{EqSets}), for $\varepsilon_1,\varepsilon_2\in \{1,-1\}$
and $i_1,i_2\in\{0,1,2\}$,
$$
N^+_{\varepsilon_1,i_1}\cap N^-_{\varepsilon_2,i_2}=\Big\{(a,b)\in \mathbb{F}_{q}^2\,|\,\big(T_0(a,b),T_0(-a,b)\big)=(\varepsilon_1\nu_{i_1},\varepsilon_2\nu_{i_2})\Big\}.
$$
Then one needs
to calculate the cardinality of $N^+_{\varepsilon_1,i_1}\cap N^-_{\varepsilon_2,i_2}$ .

It follows from Lemma \ref{Lem1-2} (ii) that for any $i_1,i_2\in\{0,1,2\}$ with $i_1\cdot i_2\neq 0$, $N^+_{\varepsilon_1,i_1}\cap N^-_{\varepsilon_2,i_2}=\emptyset$. Thus it suffices to consider the cases where $i_1\cdot i_2=0$, namely, $$(i_1,i_2)\in\{(0,0),(0,1),(0,2),(1,0),(2,0)\}.$$
Furthermore, by Lemma \ref{Lem3} (i), for any $\varepsilon_1,\varepsilon_2\in \{1,-1\}$
and $i_1,i_2\in\{0,1,2\}$, $(a,b)\in N^+_{\varepsilon_1,i_1}\cap N^-_{\varepsilon_2,i_2}$ if and only if $(-a,b)\in N^-_{\varepsilon_1,i_1}\cap N^+_{\varepsilon_2,i_2}$. Thus,
\begin{equation}\label{EqTh1-0}
  |N^+_{\varepsilon_1,i_1}\cap N^-_{\varepsilon_2,i_2}|=|N^+_{\varepsilon_2,i_2}\cap N^-_{\varepsilon_1,i_1}|.
\end{equation} Hence
we in the sequel only need to calculate $|N^+_{\varepsilon_1,i_1}\cap N^-_{\varepsilon_2,i_2}|$ for $i_1=0$ and $i_2\in \{0,1,2\}$. The cardinality
of $N^+_{\varepsilon_1,i_1}\cap N^-_{\varepsilon_2,i_2}$ for $i_1\in \{0,1,2\}$ and $i_2=0$ will be directly obtained.

Let $\lambda$ be a non-square of $\mathbb{F}_{p}^*$. Due to Lemma \ref{Lem3} (ii), we get
$$
\lambda (N^{+}_{1,0}\cap N^{-}_{1,0})=N^{+}_{-1,0}\cap N^{-}_{-1,0},\ \lambda (N^{+}_{-1,0}\cap N^{-}_{1,0})=N^{+}_{1,0}\cap N^{-}_{-1,0},
$$ which implies
\begin{equation}\label{EqTh1-1}
|N^{+}_{1,0}\cap N^{-}_{1,0}|=|N^{+}_{-1,0}\cap N^{-}_{-1,0}|,\
|N^{+}_{-1,0}\cap N^{-}_{1,0}|=|N^{+}_{1,0}\cap N^{-}_{-1,0}|.
\end{equation}
Similarly, Lemma \ref{Lem3} (iv) gives
\begin{equation}\label{EqTh1-2}
|N^{+}_{1,0}\cap N^{-}_{1,2}|=|N^{+}_{-1,0}\cap N^{-}_{-1,2}|,\
|N^{+}_{-1,0}\cap N^{-}_{1,2}|=|N^{+}_{1,0}\cap N^{-}_{-1,2}|.
\end{equation}
By Lemma \ref{Lem3} (iii), we can deduce that
\begin{equation}\label{EqTh1-3}
|N^{+}_{1,0}\cap N^{-}_{1,1}|=|N^{+}_{-1,0}\cap N^{-}_{1,1}|,\
|N^{+}_{-1,0}\cap N^{-}_{-1,1}|=|N^{+}_{1,0}\cap N^{-}_{-1,1}|.
\end{equation}
For the ease of notations,
for $i\in\{0,2\}$, denote
\begin{equation}\label{EqTh1-Notation1}
    s_{i}=|N^{+}_{1,0}\cap N^{-}_{1,i}|, \ \bar{s}_{i}=|N^{+}_{-1,0}\cap N^{-}_{1,i}|
\end{equation} and let
\begin{equation}\label{EqTh1-Notation2}
    s_1=|N^{+}_{1,0}\cap N^{-}_{1,1}|, \  \bar{s}_1=|N^{+}_{-1,0}\cap N^{-}_{-1,1}|.
\end{equation}

From (\ref{EqTh1-0})-(\ref{EqTh1-Notation2}), the reader will observe that the quantities $s_0, \overline{s}_0, s_1,\bar{s}_1,s_2,\bar{s}_2$
respectively correspond to the first item to the sixth item of the multiplicities in Table \ref{Tab2}.
Thus our next task is to determine these quantities.

By Lemma \ref{Lem3} (i), the cardinalities of
$N^{+}_{\varepsilon,i}$ and $N^{-}_{\varepsilon,i}$ are the same and they are listed in Table \ref{Tab1}. We denote $n_{\varepsilon,i}=|N^{+}_{\varepsilon,i}|=|N^{-}_{\varepsilon,i}|$.
By Lemma \ref{Lem1-2} (ii), for any $(a,b)\in \mathbb{F}_{q}\setminus\{(0,0)\}$, only when the quadratic form $Q_{a,b}(x)=ax^{p^k+1}+bx^2$ has rank $m$, the quadratic form $Q_{-a,b}(x)$ could have rank $m-2$, $m-1$. This is equivalent to saying that
 $$N^{-}_{\varepsilon,i}\subseteq (N^{+}_{1,0}\cup N^{+}_{-1,0})$$ for $\varepsilon\in\{1,-1\}$ and $i=1,2$.
Thus we have
$$
|N^{+}_{1,0}\cap N^{-}_{\varepsilon,i}|+|N^{+}_{-1,0}\cap N^{-}_{\varepsilon,i}|=|N^{-}_{\varepsilon,i}|=n_{\varepsilon,i}.
$$
This fact combined with  (\ref{EqTh1-1}), (\ref{EqTh1-2}) and (\ref{EqTh1-3}) yields the following equations
\begin{equation}\label{EqTh1-7}
  \left\{\begin{array}{l}
s_0+\bar{s}_0+s_1+\bar{s_1}+s_2+\bar{s_2}=n_{1,0}\\
s_1+s_1=n_{1,1}\\
\bar{s}_1+\bar{s}_1=n_{-1,1}\\
s_2+\bar{s}_2=n_{1,2}.
\end{array}\right.
\end{equation}

Furthermore, by the correspondences between the first six items of multiplicities in Table \ref{Tab2} and the quantities $s_0, \overline{s}_0, s_1,\bar{s}_1,s_2,\bar{s}_2$, it is easy to verify that
$$\sum\limits_{a,b\in \mathbb{F}_{q}}T_0(a,b)T_0(-a,b)=p^{2m}+2(s_0-\bar{s}_0)\nu_0^2+4(s_2-\bar{s}_2)\nu_0\nu_2$$
and
$$\sum\limits_{a,b\in \mathbb{F}_{q}}T^3_0(a,b)T_0(-a,b)=p^{4m}+2(s_0-\bar{s}_0)\nu_0^4+2(s_2-\bar{s}_2)(\nu_0^3\nu_2+\nu_0\nu_2^3).$$
Then Proposition \ref{prop2} gives two more equations
\begin{equation}\label{EqTh1-8}
  \left\{\begin{array}{l}
2(s_0-\bar{s}_0)\nu_0^2+4(s_2-\bar{s}_2)\nu_0\nu_2=0\\
2(s_0-\bar{s}_0)\nu_0^4+2(s_2-\bar{s}_2)(\nu_0^3\nu_2+\nu_0\nu_2^3)=p^{2m}(p^m-1)(p^{m}-p).
\end{array}\right.
\end{equation}
Therefore, one can obtain $s_0,\bar{s}_0,s_1,\bar{s}_1,s_2,\bar{s}_2$ by solving the systems of
equations (\ref{EqTh1-7}) and (\ref{EqTh1-8}). Then the distribution of $\big(T_0(a,b), T_0(-a,b)\big)$ is determined and listed in Table \ref{Tab2}.\hfill$\blacksquare$



\begin{table}
\begin{center}
\caption{
\small{The value distribution of $\{T(a,b): a,b\in \mathbb{F}_{q}\}$ for odd $e$}
}\label{Tab3-odd}
\begin{tabular}{|c|c|}
\hline Values & Multiplicity  \\
\hline
$0$& $(p^m-1)(p^{m}-p^{m-1}+1)$ \\  \hline
$p^{\frac{m+1}{2}}$ &$(p^m-1)(p^{m-1}+p^{\frac{m-1}{2}})/2$\\ \hline
$-p^{\frac{m+1}{2}}$ &$(p^m-1)(p^{m-1}-p^{\frac{m-1}{2}})/2$\\ \hline
$p^m$ & 1  \\
\hline
\end{tabular}
\end{center}
\end{table}

\begin{table}
\begin{center}
\caption{\small{The value distribution of $\{T(a,b): a,b\in \mathbb{F}_{q}\}$ for even $e$}}\label{Tab3}
\begin{tabular}{|c|c|}
\hline Values & Multiplicity (each) \\
\hline
$0$& $(p-1)(p^{2m}-1)/(2(p+1))$ \\  \hline
$\pm \sqrt{p^*}p^{\frac{m-1}{2}}$ & $(p^m-1)(p^{m+1}-3p^m+p+1)/(4(p-1))$ \\ \hline
$\frac{1}{2}(\pm \sqrt{p^*}+p)p^{\frac{m-1}{2}}$ &$(p^m-1)(p^{m-1}+p^{\frac{m-1}{2}})/2$\\ \hline
$\frac{1}{2}(\pm \sqrt{p^*}-p)p^{\frac{m-1}{2}}$ &$(p^m-1)(p^{m-1}-p^{\frac{m-1}{2}})/2$\\ \hline
$\pm\frac{1}{2}(1+p)\sqrt{p^*}p^{\frac{m-1}{2}}$ & ${(p^m-1)(p^{m-1}-1) }/ (p^2-1)$\\ \hline
$p^m$ & 1  \\
\hline
\end{tabular}
\end{center}
\end{table}

\begin{table}
\begin{center}
\caption{\small{The value distribution of $\{S(a,b,c): a,b\in \mathbb{F}_{q}, c\in \Omega\}$ for odd $e$}}\label{Tab4-odd}
\begin{tabular}{|c|c|}
\hline Values & Multiplicity (each) \\ \hline
$1-\frac{1}{2}(\omega^t+\omega^{-t})\pm \frac{1}{2}(\omega^t-\omega^{-t})\sqrt{p^*}p^{\frac{m-1}{2}}$ & $p^2(p^m-1)(p^m-p^{m-1}-p^{m-2}+1)/(2(p^2-1))$\\ \hline
$1-\frac{1}{2}(\omega^t+\omega^{-t})\pm \frac{1}{2}(\omega^t+\omega^{-t})p^{\frac{m+1}{2}}$ & $(p^m-1)(p^{m-1}\pm p^{\frac{m-1}{2}})/2$\\ \hline
$1-\frac{1}{2}(\omega^t+\omega^{-t})\pm \frac{1}{2}(\omega^t-\omega^{-t})\sqrt{p^*}p^{\frac{m+1}{2}}$ & $(p^m-1)(p^{m-1}-1)/(2(p^2-1))$\\ \hline
$1+\frac{1}{2}(p^m-1)(\omega^t+\omega^{-t}) $ & $1$ \\ \hline
\end{tabular}
\end{center}
\hspace{2.2cm} where $t=0,1,\cdots, p-1$.
\end{table}

\begin{table}
\begin{center}
\caption{\small{The value distribution of $\{S(a,b,c): a,b\in \mathbb{F}_{q}, c\in \Omega\}$ for even $e$}}\label{Tab4}
\begin{tabular}{|c|c|}
\hline Values & Multiplicity (each) \\
\hline
$1-\frac{1}{2}(\omega^t+\omega^{-t})\pm \frac{1}{2}(\omega^t+\omega^{-t})\sqrt{p^*}p^{\frac{m-1}{2}}$ & $(p^m-1)(p^{m+1}-3p^m+p+1)/(4(p-1))$ \\ \hline
$1-\frac{1}{2}(\omega^t+\omega^{-t})\pm \frac{1}{2}(\omega^t-\omega^{-t})\sqrt{p^*}p^{\frac{m-1}{2}}$ & $(p-1)(p^{2m}-1)/(4(p+1))$ \\ \hline
$1-\frac{1}{2}(\omega^t+\omega^{-t})+\frac{1}{2}(\pm\omega^t \sqrt{p^*}+ \omega^{-t}p)p^{\frac{m-1}{2}}$ &$(p^m-1)(p^{m-1}+p^{\frac{m-1}{2}})/2$\\ \hline
$1-\frac{1}{2}(\omega^t+\omega^{-t})+\frac{1}{2}(\pm \omega^{t}\sqrt{p^*}-\omega^{-t}p)p^{\frac{m-1}{2}}$ & $(p^m-1)(p^{m-1}- p^{\frac{m-1}{2}})/2$\\ \hline
$1-\frac{1}{2}(\omega^t+\omega^{-t})\pm\frac{1}{2}(\omega^t+\omega^{-t}p)\sqrt{p^*}p^{\frac{m-1}{2}}$ & ${(p^m-1)(p^{m-1}-1) }/ (p^2-1)$\\ \hline
$1+\frac{1}{2}(p^m-1)(\omega^t+\omega^{-t}) $ & $1$ \\
\hline
\end{tabular}
\end{center}
\hspace{2.2cm} where $t=0,1,\cdots, p-1$.
\end{table}

\smallskip

By (\ref{EqT0})-(\ref{EqS}), Tables \ref{Tab1} and \ref{Tab2},
we have the following two theorems.

\begin{thm}\label{Thm2}
Let $T(a,b)$ be the exponential sum defined  in (\ref{EqExpSumT}). For $p \equiv 3 \pmod{4}$, odd $m\geq 3$ and
an integer $e$ satisfying the Congruence Condition,
the value distribution of the multi-set
$\{T(a,b)|a,b\in \mathbb{F}_{q}\}$ is shown
  in Table \ref{Tab3-odd} if $e$ is odd and in Table \ref{Tab3} if $e$ is even.
\end{thm}

\begin{thm}\label{Thm3} Let $S(a,b,c)$ be the exponential sum defined  in (\ref{EqExpSumS}).
For $p\equiv 3 \pmod{4}$, odd $m\geq 3$ and
an integer $e$ satisfying the Congruence Condition, the value distribution of the multi-set
$\{S(a,b,c)|a,b\in \mathbb{F}_{q}, c\in \Omega\}$ is shown
  in Table \ref{Tab4-odd} if $e$ is odd and in Table \ref{Tab4} if $e$ is even..
\end{thm}

\section{Weight distribution of $\mathcal{C}_{(1,e,s)}$}\label{Sec4}

In this section, for odd $m\geq 3$ and $p\equiv 3 \pmod{4}$, we study the weight distribution of the cyclic codes $\mathcal{C}_{(1,e,s)}$ for integers $e$ satisfying the Congruence Condition, i.e., there
exist integers $k$ coprime to $m$ and $\tau\in \mathbb{Z}_m$ such that
$
(p^k+1)e\equiv 2p^{\tau} \pmod{p^m-1}.
$

\begin{thm}\label{Thm5}
 Let $p\equiv 3 \pmod{4}$, $s=(p^m-1)/2$, $m\geq 3$ be an odd integer and $e$ be an integer satisfying the Congruence Condition. Then the weight distribution $\{A_0,A_1,\cdots,A_{q-1}\}$ of the cyclic code $\mathcal{C}_{(1,e,s)}$ is shown as follows.

(i) If $e$ is odd, $A_i=0$ expect the following
 $$
\begin{array}{cc}
i & A_i
\\
p^{m-1}(p-1)-p^{\frac{m-1}{2}}-1, & \frac{(p^m-1)(p^{m+2}-p^m-p^{m-1}-p^{\frac{m+3}{2}}+p^{\frac{m-1}{2}}+p^2)}{2(p+1)}\\
p^{m-1}(p-1)+ p^{\frac{m-1}{2}}-1, & \frac{(p^m-1)(p^{m+2}-p^m-p^{m-1}+p^{\frac{m+3}{2}}-p^{\frac{m-1}{2}}+p^2)}{2(p+1)}\\
p^{m-1}(p-1)-p^{\frac{m+1}{2}}-1, & \frac{(p^m-1)(p^{m-1}-1)}{2(p+1)}\\
p^{m-1}(p-1)+p^{\frac{m+1}{2}}-1, & \frac{(p^m-1)(p^{m-1}-1)}{2(p+1)}\\
p^{m-1}(p-1)-(p-1)p^{\frac{m-1}{2}}, & \frac{(p^m-1)(p^{m-1}+p^{\frac{m-1}{2}})}{2}\\
p^{m-1}(p-1)+(p-1)p^{\frac{m-1}{2}}, & \frac{(p^m-1)(p^{m-1}-p^{\frac{m-1}{2}})}{2}\\
p^{m-1}(p-1), &(p^m-1)(p^m-p^{m-1}+1)\\
p^{m}- 1, & p-1\\
0,& 1.
 \end{array}
 $$
(ii) If $e$ is even,  $A_i=0$ expect the following
 $$
\begin{array}{cc}
i & A_i
\\
p^{m-1}(p-1)-p^{\frac{m-1}{2}}-1, & \frac{(p^m-1)(p^{m+1}+p^m+2p^{m-1}-2p^{\frac{m+1}{2}}-2p^{\frac{m-1}{2}}+p-1)(p-1)}{4(p+1)}\\
p^{m-1}(p-1)+p^{\frac{m-1}{2}}-1, & \frac{(p^m-1)(p^{m+1}+p^m+2p^{m-1}+2p^{\frac{m+1}{2}}+2p^{\frac{m-1}{2}}+p-1)(p-1)}{4(p+1)}\\
p^{m-1}(p-1)-\frac{1}{2}(p-1)p^{\frac{m-1}{2}}-1, & \frac{(p^m-1)(p^{m-1}-1)}{p+1}\\
p^{m-1}(p-1)+\frac{1}{2}(p-1)p^{\frac{m-1}{2}}-1, & \frac{(p^m-1)(p^{m-1}-1)}{p+1}\\
p^{m-1}(p-1)-\frac{1}{2}(p-1)p^{\frac{m-1}{2}}, & (p^m-1)(p^{m-1}+p^{\frac{m-1}{2}})\\
p^{m-1}(p-1)+\frac{1}{2}(p-1)p^{\frac{m-1}{2}}, & (p^m-1)(p^{m-1}-p^{\frac{m-1}{2}}) \\
p^{m-1}(p-1)-1, & \frac{(p^m-1)(p^{m+1}-p^m-2p^{m-1}+p+1)}{2}\\
p^{m-1}(p-1), & (p^m-1)(p^{m}+1-2p^{m-1})\\
p^m-1, & p-1\\
0,& 1.
 \end{array}
 $$
\end{thm}

\noindent\textbf{Proof.}
By (\ref{EqHammingWt}),
the Hamming weight of any nonzero codeword $\mathbf{c}=(c_0,c_1,\cdots,c_{q-2})\in \mathcal{C}_{(1,e,s)}$ is
\begin{equation}\label{Eq Thm5 0}
w_H(\mathbf{c})=p^{m-1}(p-1)-\frac{1}{p}\Delta(a,b,c),
\end{equation} where
 $$
 \Delta(a,b,c)=\sum\limits_{y\in \mathbb{F}_{p}^*}\sum\limits_{x\in \mathbb{F}_{q}}\omega^{y\Tr(ax+bx^e+cx^s)}.
 $$ It suffices to determine the value distribution of the exponential sum $ \Delta(a,b,c)$.

Let $\Tr(c)=t$ and $Q_{a,b}(x)=\Tr(ax^{p^k+1}+bx^2)$.
The exponential sum $ \Delta(a,b,c)$ is investigated according to the parity of the integer $e$ in the following.

When $e$ is an odd integer satisfying the Congruence Condition,  one has
\begin{equation}\label{EqThm5-0}
    \begin{array}{rcl}
\Delta(a,b,c)&=&\frac{1}{2}\sum\limits_{y\in \mathbb{F}_{p}^*}\Big(2+\sum\limits_{x\in \mathbb{F}_{q}^*}\omega^{y\Tr(ax^{p^k+1}+bx^2+c)}+\sum\limits_{x\in \mathbb{F}_{q}^*}\omega^{y\Tr(-ax^{p^k+1}-bx^2-c)}\Big)
\\&=&\frac{1}{2}\sum\limits_{y\in \mathbb{F}_{p}^*}\Big(2+\sum\limits_{x\in \mathbb{F}_{q}^*}\omega^{y\Tr(ax^{p^k+1}+bx^2+c)}+\sum\limits_{x\in \mathbb{F}_{q}^*}\omega^{y\Tr(ax^{p^k+1}+bx^2+c)}\Big)
\\&=&\sum\limits_{y\in \mathbb{F}_{p}^*}\Big(1+\sum\limits_{x\in \mathbb{F}_{q}^*}\omega^{y\Tr(ax^{p^k+1}+bx^2+c)}\Big)
\\&=&\sum\limits_{y\in \mathbb{F}_{p}^*}\Big(1-\omega^{yt}
+\sum\limits_{x\in \mathbb{F}_{q}}\omega^{yQ_{a,b}(x)+yt}\Big)
\\&=&\sum\limits_{y\in \mathbb{F}_{p}^*}\Big(1-\omega^{yt}
+\omega^{yt}\big(\frac{y^{r}}{p}\big)\sum\limits_{x\in \mathbb{F}_{q}}\omega^{Q_{a,b}(x)}\Big),
\end{array}
\end{equation}
 where $r$ is the rank of $Q(a,b)$  and
 the last equality sign comes from Lemma \ref{Lem1}.

 {\em Case I:} $t=0$. In this case,
$$\begin{array}{c}
\Delta(a,b,c)=\sum\limits_{y\in \mathbb{F}_{p}^*}\big(\frac{y^{r}}{p}\big)\sum\limits_{x\in \mathbb{F}_{q}}\omega^{Q_{a,b}(x)}.
\end{array}
$$
Thus,
it follows from Table \ref{Tab1} that
\begin{equation}\label{EqThm5-1-odd}
\Delta(a,b,c) =\left\{\begin{array}{cl}
(p-1)p^m, & \text{once}\\
(p-1)p^{\frac{m+1}{2}}, & \text{for $(p^m-1)(p^{m-1}+p^{\frac{m-1}{2}})/2$ times}\\
-(p-1)p^{\frac{m+1}{2}}, & \text{for $(p^m-1)(p^{m-1}-p^{\frac{m-1}{2}})/2$ times}\\
0, & \text{for $(p^m-1)(p^m-p^{m-1}+1)$ times.}
\end{array}
\right.
\end{equation}

{\em Case II:} $t\neq 0$. It is easily seen that
$$
\sum\limits_{y\in \mathbb{F}_{p}^*}\omega^{yt}=-1
$$
and
$$
\sum\limits_{y\in \mathbb{F}_{p}^*}\omega^{yt}\big(\frac{y}{p}\big)=\big(\frac{t}{p}\big)\sum\limits_{y\in \mathbb{F}_{p}^*}\omega^{ty}\big(\frac{ty}{p}\big)=\big(\frac{t}{p}\big)\sqrt{p^{*}}.
$$ Then,
$$
\Delta(a,b,c)=p+\sum\limits_{y\in \mathbb{F}_{p}^*}\omega^{yt}\big(\frac{y^{r}}{p}\big)\sum\limits_{x\in \mathbb{F}_{q}}\omega^{Q_{a,b}(x)}.
$$ Furthermore, by Table \ref{Tab1} we have
\begin{equation}\label{EqThm5-2-odd}
\Delta(a,b,c) =\left\{\begin{array}{cl}
p-p^m, & \text{for $p-1$ time}\\
p+ p^{\frac{m+1}{2}}, & \text{for $\frac{(p^m-1)(p^{m+2}-p^m-p^{m-1}-p^{\frac{m+3}{2}}+p^{\frac{m-1}{2}}+p^2)}{2(p+1)}$ times}\\
p- p^{\frac{m+1}{2}}, & \text{for $\frac{(p^m-1)(p^{m+2}-p^m-p^{m-1}+p^{\frac{m+3}{2}}-p^{\frac{m-1}{2}}+p^2)}{2(p+1)}$ times}\\
p\pm p^{\frac{m+3}{2}}, & \text{for $\frac{(p^m-1)(p^{m-1}-1)}{2(p+1)}$ times.}
\end{array}
\right.
\end{equation}
The weight distribution of $\mathcal{C}_{(1,e,s)}$ for odd $e$ can be derived from (\ref{Eq Thm5 0}), (\ref{EqThm5-1-odd}) and (\ref{EqThm5-2-odd}).

When  $e$ is an even integer satisfying the Congruence Condition,
\begin{equation}\label{EqThm5-0-even}
    \begin{array}{rcl}
\Delta(a,b,c)&=&\frac{1}{2}\sum\limits_{y\in \mathbb{F}_{p}^*}\Big(2+\sum\limits_{x\in \mathbb{F}_{q}^*}\omega^{y(Q_{a,b}(x)+t)}+\sum\limits_{x\in \mathbb{F}_{q}^*}\omega^{y(Q_{-a,b}(x)-t)}\Big)
\\&=&\frac{1}{2}\sum\limits_{y\in \mathbb{F}_{p}^*}\Big(2-(\omega^{yt}+\omega^{-yt})+\sum\limits_{x\in \mathbb{F}_{q}}\omega^{yQ_{a,b}(x)+yt}+\sum\limits_{x\in \mathbb{F}_{q}}\omega^{yQ_{-a,b}(x)-yt}\Big)
\\&=&\frac{1}{2}\sum\limits_{y\in \mathbb{F}_{p}^*}\Big(2-(\omega^{yt}+\omega^{-yt})+\omega^{yt}\big(\frac{y^{r}}{p}\big)\sum\limits_{x\in \mathbb{F}_{q}}\omega^{Q_{a,b}(x)}+\omega^{-yt}\big(\frac{y^{r'}}{p}\big)\sum\limits_{x\in \mathbb{F}_{q}}\omega^{Q_{-a,b}(x)}\Big)
\end{array}
\end{equation}
 where $r, r'$ are the rank of $Q(a,b)$ and $Q(-a,b)$ respectively, and
 the last equality sign comes from Lemma \ref{Lem1}. It follows from Lemma \ref{Lem1-2} that
\begin{equation}\label{Eq Thm5-r}
(r,r')\in\{(m,m),(m,m-1),(m,m-2),(m-1,m),(m-2,m)\}.
\end{equation}

{\em Case I:} $t=0$. In this case,
$$\begin{array}{c}
\Delta(a,b,c)=\frac{1}{2}\Big(\sum\limits_{y\in \mathbb{F}_{p}^*}\big(\frac{y^{r}}{p}\big)\sum\limits_{x\in \mathbb{F}_{q}^*}\omega^{Q_{a,b}(x)}+\sum\limits_{y\in \mathbb{F}_{p}^*}\big(\frac{y^{r'}}{p}\big)\sum\limits_{x\in \mathbb{F}_{q}^*}\omega^{Q_{-a,b}(x)}\Big).
\end{array}
$$
By a similar analysis as in the case of odd $e$, one deduces
\begin{equation}\label{EqThm5-1}
\Delta(a,b,c) =\left\{\begin{array}{cl}
(p-1)p^m, & \text{once}\\
\frac{1}{2}(p-1)p^{\frac{m-1}{2}}, & \text{for $(p^m-1)(p^{m-1}+p^{\frac{m-1}{2}})$ times}\\
-\frac{1}{2}(p-1)p^{\frac{m-1}{2}}, & \text{for $(p^m-1)(p^{m-1}-p^{\frac{m-1}{2}})$ times}\\
0, & \text{for $(p^m-1)(p^{m}-2p^{m-1}+1)$ times.}
\end{array}
\right.
\end{equation}

{\em Case II:} $t\neq 0$. Since $
\sum_{y\in \mathbb{F}_{p}^*}\omega^{yt}=\sum_{y\in \mathbb{F}_{p}^*}\omega^{-yt}=-1
$, one has
$$\begin{array}{c}
\Delta(a,b,c)=p+\frac{1}{2}\Big(\sum\limits_{y\in \mathbb{F}_{p}^*}\omega^{yt}\big(\frac{y^{r}}{p}\big)\sum\limits_{x\in \mathbb{F}_{q}}\omega^{Q_{a,b}(x)}+\sum\limits_{y\in \mathbb{F}_{p}^*}\omega^{-yt}\big(\frac{y^{r'}}{p}\big)\sum\limits_{x\in \mathbb{F}_{q}}\omega^{Q_{-a,b}(x)}\Big).
\end{array}
$$Furthermore, the fact $\sum_{y\in \mathbb{F}_{p}^*}\omega^{yt}\big(\frac{y}{p}\big)=\big(\frac{t}{p}\big)\sqrt{p^{*}}$ and the possible values of $(r,r')$ in (\ref{Eq Thm5-r})
 imply
\begin{equation}\label{EqThm5-2}
\Delta(a,b,c)
=\left\{\begin{array}{cl}
  p+\frac{1}{2}\Big(\big(\frac{t}{p}\big)\sqrt{p^*}\sum\limits_{x\in \mathbb{F}_{q}}\omega^{Q_{a,b}(x)}-\sum\limits_{x\in \mathbb{F}_{q}}\omega^{Q_{-a,b}(x)}\Big), & \text{if $(r,r')=(m,m-1)$}\\
 p+\frac{1}{2}\Big(-\sum\limits_{x\in \mathbb{F}_{q}}\omega^{Q_{a,b}(x)}-\big(\frac{t}{p}\big)\sqrt{p^*}\sum\limits_{x\in \mathbb{F}_{q}}\omega^{Q_{-a,b}(x)}\Big), & \text{if $(r,r')=(m-1,m)$}\\
  p+\frac{1}{2}\big(\frac{t}{p}\big)\sqrt{p^*}\Big(\sum\limits_{x\in \mathbb{F}_{q}}\omega^{Q_{a,b}(x)}-\sum\limits_{x\in \mathbb{F}_{q}}\omega^{Q_{-a,b}(x)}\Big), & \text{otherwise.}\\
\end{array}
\right.
\end{equation}
Recall that $T_0(a,b)=\sum\limits_{x\in \mathbb{F}_{q}}\omega^{Q_{a,b}(x)}$.
Thus, by the distribution of $\big(T_0(a,b), T_0(-a,b)\big)$ in Table \ref{Tab2}, we deduce that
\begin{equation}\label{EqThm5-3}
\Delta(a,b,c)
=\left\{\begin{array}{cl}
 p+p^{\frac{m+1}{2}}, & \text{for $\frac{(p^m-1)(p^{m+1}+p^m+2p^{m-1}-2p^{\frac{m+1}{2}}-2p^{\frac{m-1}{2}}+p-1)(p-1)}{4(p+1)}$ times}\\
  p-p^{\frac{m+1}{2}}, & \text{for $\frac{(p^m-1)(p^{m+1}+p^m+2p^{m-1}+2p^{\frac{m+1}{2}}+2p^{\frac{m-1}{2}}+p-1)(p-1)}{4(p+1)}$ times}\\
  p+\frac{1}{2}(p-1)p^{\frac{m+1}{2}}, & \text{for $\frac{(p^m-1)(p^{m-1}-1)}{p+1}$ times}\\
  p-\frac{1}{2}(p-1)p^{\frac{m+1}{2}}, & \text{for $\frac{(p^m-1)(p^{m-1}-1)}{p+1}$ times}\\
 p, & \text{for $\frac{(p^m-1)(p^{m+1}-p^m-2p^{m-1}+p+1)}{2}$ times}\\
 p-q & \text{for $(p-1)$ times}.
\end{array}
\right.
\end{equation} By (\ref{Eq Thm5 0}), (\ref{EqThm5-1}) and (\ref{EqThm5-3}), we deduce the weight
distribution of $\mathcal{C}_{(1,e,s)}$ for even $e$. \hfill$\blacksquare$

\medskip

Theorem \ref{Thm5} settles the weight distribution of $\mathcal{C}_{(1,e,s)}$ for $p\equiv 3 \pmod{4}$
and integers $e$ satisfying the Congruence Condition. For the special case $p=3$, as pointed in the end of Section \ref{Sec2-1}, the
following APN exponents
\begin{enumerate}
  \item[1)] $e=\frac{3^m+1}{4}+\frac{3^m-1}{2}$;
  \item[2)] $e=3^{(m+1)/2}-1$; and
  \item[3)] $e=(3^{(m+1)/4}-1)(3^{(m+1)/2}+1)$ for $m\equiv 3 \pmod{4}$
\end{enumerate}
satisfy the Congruence Condition and generate the optimal cyclic codes $\mathcal{C}_{(1,e,s)}^{\bot}$ in \cite{LLHDT}.
Thus we have the following corollary.
\begin{cor}\label{Cor2}
 For the three APN exponents
 \begin{enumerate}
  \item[1)] $e=\frac{3^m+1}{4}+\frac{3^m-1}{2}$;
  \item[2)] $e=3^{(m+1)/2}-1$; and
  \item[3)] $e=(3^{(m+1)/4}-1)(3^{(m+1)/2}+1)$ for $m\equiv 3  \pmod{4}$,
\end{enumerate}
the weight distribution of the cyclic code $\mathcal{C}_{(1,e,s)}$ is given as in Table \ref{Tab6}.
\end{cor}

%

\begin{table}
\begin{center}
\caption{\small{The weight distribution of $\mathcal{C}_{(1,e,s)}$ for three APN exponents $e$ and $s=(3^m-1)/2$}}\label{Tab6}
\begin{tabular}{|c|c|}
\hline Hamming weight & Multiplicity \\
\hline
$2\cdot 3^{m-1}-3^{\frac{m-1}{2}}-1$ & $ (3^m-1) (2\cdot 3^{m-1}-3^{\frac{m-1}{2}})$\\ \hline
$2\cdot 3^{m-1}+3^{\frac{m-1}{2}}-1$ & $ (3^m-1) (2\cdot 3^{m-1}+3^{\frac{m-1}{2}})$ \\ \hline
$2\cdot 3^{m-1}-3^{\frac{m-1}{2}}$ & $(3^m-1)(3^{m-1}+3^{\frac{m-1}{2}})$\\ \hline
$2\cdot 3^{m-1}+3^{\frac{m-1}{2}}$ & $(3^m-1)(3^{m-1}-3^{\frac{m-1}{2}})$ \\ \hline
$2\cdot 3^{m-1}-1$ & $2(3^m-1) (3^{m-1}+1)$\\ \hline
$2\cdot 3^{m-1}$ & $(3^m-1)(3^{m-1}+1)$\\ \hline
$3^m-1$ & $2$\\ \hline
$0$& $1$\\ \hline
\end{tabular}
\end{center}
\end{table}

\section{Summary and concluding remarks}\label{sec-last}

One major contribution of this paper is the development of Theorem \ref{Thm1}, which not only unifies
the weight distributions of the thirteen classes of cyclic codes documented in \cite{CKNC,Zhou1} and \cite{Zhou2},
but also settles the weight distribution of many new three-weight codes $\mathcal{C}_{(1,e)}$ with two
zeros. In many cases, the duals of these three-weight codes $\mathcal{C}_{(1,e)}$ are optimal
\cite{Ding-Helleseth,Zhou1,Zhou2}.

Another major contribution of this paper is the settlement of the weight distributions of a class
of cyclic codes $\mathcal{C}_{(1,e,s)}$ with three zeros, which is documented in Theorem \ref{Thm5}.
In many cases the duals of the codes $\mathcal{C}_{(1,e,s)}$ are also optimal \cite{LLHDT}. Our technique
in proving Theorem \ref{Thm5} is similar to that employed in \cite{ZDLZ}.

\section{The Appendix}

\subsection*{Proof of Proposition \ref{prop2-1}:}
Note that the equation $x^{p^k+1}+y^{p^k+1}+z^{p^k+1}-w^{p^k+1}=0$ is equivalent to $x^{p^{m-k}+1}+y^{p^{m-k}+1}+z^{p^{m-k}+1}-w^{p^{m-k}+1}=0$.
Since $m$ is odd,  we may assume $k$ is even in the sequel (otherwise replace $k$ with $m-k$).

 For any $(\alpha,\beta)\in \mathbb{F}_{p^{m}}^2$, let $N(\alpha, \beta)$ denote the number of solutions of the
following system of equations
$$
  \left\{
    \begin{array}{l}
      x^2+y^2=\alpha\\
      x^{p^k+1}+y^{p^k+1}=\beta\\
      z^2+w^2=-\alpha\\
      z^{p^k+1}-w^{p^k+1}=-\beta.
    \end{array}
  \right.
$$Then,
\begin{equation}\label{EqM4-N}\mathcal{N}_4=\sum\limits_{\alpha,\beta\in \mathbb{F}_{p^m}}N(\alpha, \beta).\end{equation}

Given $(\alpha,\beta)\in \mathbb{F}_{p^m}^2$, we will study  the
following systems of equations:
\begin{equation}\label{EqM4-PartI}
  \left\{
    \begin{array}{l}
      x^2+y^2=\alpha\\
      x^{p^k+1}+y^{p^k+1}=\beta\\
    \end{array}
  \right.
\end{equation} and
\begin{equation}\label{EqM4-PartII}
 \left\{
    \begin{array}{l}
      z^2+w^2=-\alpha\\
      z^{p^k+1}-w^{p^k+1}=-\beta.
 \end{array}
  \right.
\end{equation}
Denote by $N_1(\alpha, \beta)$ and $N_2(\alpha, \beta)$ the numbers of
solutions of (\ref{EqM4-PartI}) and (\ref{EqM4-PartII}),
respectively. Then $N(\alpha, \beta)=N_1(\alpha, \beta)*N_2(\alpha,
\beta)$ for any $(\alpha, \beta)\in \mathbb{F}_{p^m}^2$.

We first investigate the possible values of $N(\alpha,\beta)$ for
$p\equiv 3 \pmod{4}$.
The investigation is divided into two subcases: $\alpha\beta=0$ and
$\alpha\beta\neq 0$.

\emph{Case I: $\alpha\beta=0$}. For (\ref{EqM4-PartI}), if
$\alpha=0$, then $x=y=0$ since $-1$ is a non-square in
$\mathbb{F}_{p^m}$. Thus (\ref{EqM4-PartI}) has a solution if and only if
$\beta=0$. If $\beta=0$, $x^{2(p^k+1)}=y^{2(p^k+1)}$ together with
$\gcd(2(p^k+1),p^m-1)=2$ implies $x^2=y^2$, then $x=y=0$ since
$x^{p^k+1}=-y^{p^k+1}$. Thus (\ref{EqM4-PartI}) has a solution if and only if
$\alpha=0$. Therefore, (\ref{EqM4-PartI}) has only one solution for
$(\alpha,\beta)=(0,0)$ and has no solution in other cases.  In addition, for $(\alpha,\beta)=(0,0)$, (\ref{EqM4-PartII}) becomes $z^2+w^2=0$
and $z^2-w^2=0$. Thus it has exactly one solution $(0,0)$ as well.


From the above analysis we deduce that when $\alpha\beta=0$,
\begin{equation}\label{EqM4-N12CaseI}
  N(\alpha,\beta)=\left\{
                    \begin{array}{ll}
                      1, & \hbox{if $(\alpha,\beta)=(0,0)$} \\
                      0, & \hbox{otherwise.}
                    \end{array}
                  \right.
\end{equation}

\emph{Case II: $\alpha\beta\neq 0$}. We first consider the possible
values of $N_1(\alpha,\beta)$.

Let $s,t\in \mathbb{F}_{p^{2m}}^*$ such that $s^2=\alpha$, $t^2=-1$.
It is easy to verify that all solutions $x,y\in
\mathbb{F}_{p^{2m}}^*$ of $x^2+y^2=\alpha$ have the form
\begin{equation}\label{EqM4a-1}
  x=\frac{1}{2}s(\theta+\theta^{-1}),\ \ y=\frac{1}{2}st(\theta-\theta^{-1}), \ \  \theta\in \mathbb{F}_{p^{2m}}^*.
\end{equation}

The above representations of $x$, $y$ will be utilized to analyze
the solutions of (\ref{EqM4-PartI}) in $\mathbb{F}_{p^m}$.

For the first equation of (\ref{EqM4-PartI}), by applying the fact
$x^{p^m}=x$ to (\ref{EqM4a-1}), we get
$s^{p^m-1}(\theta+\theta^{-1})^{p^m}=(\theta+\theta^{-1})$, which is
equivalent to
$$\big(\theta^{p^m+1}-\alpha^{(p^m-1)/2}\big)\big(\theta^{p^m-1}-\alpha^{(p^m-1)/2}\big)=0.$$
Similarly, the fact $y^{p^m}=y$ gives
$$\big(\theta^{p^m+1}-\alpha^{(p^m-1)/2}\big)\big(\theta^{p^m-1}+\alpha^{(p^m-1)/2}\big)=0.$$
By combining these two equations, we deduce
\begin{equation}\label{EqM4a-2}
 \theta^{p^m+1}=\alpha^{(p^m-1)/2}.
\end{equation}

It follows from the second equation of (\ref{EqM4-PartI}) and
(\ref{EqM4a-1}) that
$$
\big(\frac{1}{2}s(\theta+\theta^{-1})\big)^{p^k+1}+\big(\frac{1}{2}st(\theta-\theta^{-1})\big)^{p^k+1}
=\frac{1}{2}\alpha^{(p^k+1)/2}(\theta^{p^k-1}+\theta^{1-p^k})=\beta.
$$
Thus, if we take $\tau_1=\theta^{p^k-1}$ and
$\beta_1=2\alpha^{-(p^k+1)/2}\beta$, then $\tau_1, \tau_1^{-1}$
are the two solutions of the following equation
\begin{equation}\label{EqM4a-3}
\tau^2-\beta_1\tau+1=0.
\end{equation}

By (\ref{EqM4a-2}) and (\ref{EqM4a-3}), we have
\begin{equation}\label{EqM4b-4}
    \theta^{p^m+1}=\alpha^{(p^m-1)/2}, \quad
    \theta^{p^k-1}=\tau_1
\end{equation}
and
\begin{equation}\label{EqM4b-5}
    \theta^{p^m+1}=\alpha^{(p^m-1)/2},   \quad  \theta^{p^k-1}=\tau_1^{-1}.
\end{equation}

If $\beta_1=2$, then $\tau_1=\tau^{-1}_1=\theta^{p^k-1}=1$ and
(\ref{EqM4b-5}) is the same as (\ref{EqM4b-4}). Note that $\gcd(2(p^m+1), p^k-1)=2(p+1)$ when $k$ is even and $\gcd(m,k)=1$.  Then
$\theta^{2(p^m+1)}=\alpha^{(p^m-1)}=1$ together with $\theta^{p^k-1}=1$ gives $\theta^{2(p+1)}=1$. Furthermore, since $m$ is odd and $\alpha^{(p^m-1)/2}=\pm 1$, one gets
$\theta^{p+1}=\alpha^{(p^m-1)/2}$, which indicates that
(\ref{EqM4b-4}) has exactly $p+1$ solutions in $\mathbb{F}_{p^{2m}}$.

 If $\beta_1=-2$, then $\tau_1=\tau_1^{-1}=\theta^{p^k-1}=-1$ and (\ref{EqM4b-5}) is the same as  (\ref{EqM4b-4}). The
 fact $\theta^{2(p^m+1)}=\theta^{2(p^k-1)}=1$ suggests $\theta^{\gcd(2(p^m+1),\,2(p^k-1))}=\theta^{2(p+1)}=1$, which is in contradiction with $\theta^{p^k-1}=-1$ since $2(p+1)|(p^k-1)$. Thus, (\ref{EqM4b-4}) has no solution in this case.

If $\beta_1\neq \pm 2$, then $\tau_1\neq\tau_1^{-1}$. It is readily
seen that $\theta$ is a solution of (\ref{EqM4b-4}) if and only if
$\theta^{-1}$ is a solution of (\ref{EqM4b-5}). Suppose $\theta_1$
and $\theta_2$ are two solutions of (\ref{EqM4b-4}). Then
$(\theta_1/\theta_2)^{p^m+1}=(\theta_1/\theta_2)^{p^k-1}=1$, and
this implies $(\theta_1/\theta_2)^{p+1}=1$ since
$\gcd(p^m+1,p^k-1)=p+1$. As a result, if (\ref{EqM4b-4}) has a
solution $\theta$, all solutions of (\ref{EqM4b-4}) can be
represented as $\mu\theta$, and all solutions of (\ref{EqM4b-5}) can
be represented as $\mu\theta^{-1}$, where $\mu\in \mathbb{F}_{p^{2m}}$ and $\mu^{p+1}=1$.
Therefore  for (\ref{EqM4b-4}) and (\ref{EqM4b-5}), either each of
them has exactly $p+1$ solutions or none of them has a solution.

In summary, for $\alpha\beta\neq 0$,
\begin{equation}\label{EqM4bb-N1}
    N_1(\alpha,\beta)=\left\{
                        \begin{array}{ll}
                          p+1, & \hbox{if $\beta=\alpha^{(p^k+1)/2}$,} \\
                          0, & \hbox{if $\beta=-\alpha^{(p^k+1)/2}$,} \\
                          0 \text{ or } 2(p+1), & \hbox{otherwise.}
                        \end{array}
                      \right.
\end{equation}

Now we turn to analyze the possible values of $N_2(\alpha,\beta)$.
The analysis proceeds in a similar fashion to that for
$N_1(\alpha,\beta)$.

Recall that $s, t\in \mathbb{F}_{p^{2m}}^*$ with $s^2=\alpha$ and
$t^2=-1$. Thus all solutions of $z^2+w^2=-\alpha$ can be represented
as
\begin{equation}\label{EqM4c-1}
  z= \frac{1}{2}st(\vartheta+\vartheta^{-1}), \ \
  w=\frac{1}{2}s(\vartheta-\vartheta^{-1}), \ \
  \vartheta\in \mathbb{F}_{p^{2m}}^*.
\end{equation}
For the first equation of (\ref{EqM4-PartII}), the facts $z^{p^m}=z$
and $w^{p^m}=w$ imply
\begin{equation}\label{EqM4c-2}
\vartheta^{p^m+1}=-\alpha^{(p^m-1)/2},
\end{equation} and
the second equation of (\ref{EqM4-PartII}) together with
(\ref{EqM4c-1}) yields
$$
\frac{1}{2}\alpha^{(p^k+1)/2}
(\vartheta^{p^{k}+1}+\vartheta^{-(p^{k}+1)})=\beta.
$$ Assume $\tau_2=\vartheta^{p^{k}+1}$ and $\beta_1=2\alpha^{-(p^k+1)/2}\beta$. Then $\tau_2, \tau_2^{-1}$ are the two solutions of
\begin{equation}\label{EqM4c-3}
    \tau^2-\beta_1\tau+1=0.
\end{equation}
This equation is the same as Equation
(\ref{EqM4a-3}). Thus  $\tau_2\in \{\tau_1, \tau_1^{-1}\}$.

By (\ref{EqM4c-2}) and (\ref{EqM4c-3}), we get the following
equations:
\begin{equation}\label{EqM4d-4}
   \vartheta^{p^m+1}=-\alpha^{(p^m-1)/2},\quad \vartheta^{p^k+1}=\tau_2
\end{equation} and
\begin{equation}\label{EqM4d-5}
  \vartheta^{p^m+1}=-\alpha^{(p^m-1)/2},\quad \vartheta^{p^k+1}=\tau_2^{-1}.
\end{equation}

If $\beta_1=2$, then $\tau_2=\tau_2^{-1}=\vartheta^{p^k+1}=1$ and
(\ref{EqM4d-5}) is the same as (\ref{EqM4d-4}). Since
$\gcd(2(p^m+1),p^k+1)=2$, one deduces $\vartheta^{2}=1$. Therefore,
(\ref{EqM4d-4}) has exactly $2$ solutions $\vartheta=\pm 1$ if
$\alpha$ is a non-square and has no solution otherwise.

 If $\beta_1=-2$, then $\tau_2=\tau_2^{-1}=\vartheta^{p^k+1}=-1$ and (\ref{EqM4d-5}) is the same as (\ref{EqM4d-4}). The
 fact $\gcd(2(p^m+1),2(p^k+1))=4$ suggests $\vartheta^{4}=1$, and then $\vartheta^{2}=-1$.
Thus, (\ref{EqM4d-4}) has exactly $2$ solutions $\vartheta=\pm t$ if
$\alpha$ is a non-square, where $t^2=-1$, and has no solution
otherwise.

If $\beta_1\neq \pm 2$, then $\tau_2\neq\tau_2^{-1}$. Note that
$\vartheta$ is a solution of (\ref{EqM4d-4}) if and only if
$\vartheta^{-1}$ is a solution of (\ref{EqM4d-5}). Suppose
$\vartheta_1$ and $\vartheta_2$ are two solutions of
(\ref{EqM4d-4}). Then
$(\vartheta_1/\vartheta_2)^{p^m+1}=(\vartheta_1/\vartheta_2)^{p^k+1}=1$,
and this implies $(\vartheta_1/\vartheta_2)^{2}=1$. Consequently,
for (\ref{EqM4d-4}) and (\ref{EqM4d-5}), either they respectively
have solutions $\pm \vartheta$ and $\pm \vartheta^{-1}$, or none of
them has a solution.

Summarizing up, for $\alpha\beta\neq 0$, we have
\begin{equation}\label{EqM4d-N2}
    N_2(\alpha,\beta)=\left\{
                        \begin{array}{ll}
                          2, & \hbox{if $\beta=\pm\alpha^{(p^k+1)/2}$ and $\alpha^{(p^m-1)/2}=-1$,} \\
                          0, & \hbox{if $\beta=\pm\alpha^{(p^k+1)/2}$ and $\alpha^{(p^m-1)/2}=1$,} \\
                          0 \text{ or } 4, & \hbox{otherwise.}
                        \end{array}
                      \right.
\end{equation}

By (\ref{EqM4bb-N1}) and (\ref{EqM4d-N2}), for $\alpha\in
\mathbb{F}_{p^m}^*$ and $\beta\in
\{\alpha^{(p^k+1)/2},-\alpha^{(p^k+1)/2}\}$,
\begin{equation}\label{EqM4-NCaseII}
    N(\alpha,\beta)=\left\{
                        \begin{array}{ll}
                          2(p+1), & \hbox{if $\alpha$ is a non-square and $\beta=\alpha^{(p^k+1)/2}$,} \\
                          0 , & \hbox{otherwise}.
                        \end{array}
                      \right.
\end{equation}
To complete the proof, the next task is to consider the possible
values of $N(\alpha,\beta)$ for $\alpha\in \mathbb{F}_{p^m}^*$ and
$\beta\in \mathbb{F}_{p^m}^*\setminus \{\pm\alpha^{(p^k+1)/2}\}$.
Thus we turn back to Equations (\ref{EqM4b-4}), (\ref{EqM4b-5}),
(\ref{EqM4d-4}) and (\ref{EqM4d-5}) and gather them together as
below
\begin{equation}\label{EqM4Equtions}
\left\{
  \begin{array}{ll}
    \theta^{p^m+1}=\alpha^{(p^m-1)/2}, &\theta^{(p^k-1)}=\tau_1, \text{ or } \tau_1^{-1}, \\
    \vartheta^{p^m+1}=-\alpha^{(p^m-1)/2},& \vartheta^{(p^k+1)}=\tau_1,\text{ or } \tau_1^{-1}
  \end{array}
\right.
\end{equation}
since $\tau_2\in\{\tau_1,\tau_1^{-1}\}$. For a fixed $\alpha\in
\mathbb{F}_{p^m}^*$, define
\begin{equation}\label{EqM4-Sets}\begin{array}{l}
S_1(\alpha)=\Big\{\beta\in \mathbb{F}_{p^m}^*\setminus \{\pm\alpha^{(p^k+1)/2}\}\,|\,\text{both (\ref{EqM4b-4}) and (\ref{EqM4b-5}) have $p+1$ solutions} \Big\},\\
S_2(\alpha)=\Big\{\beta\in \mathbb{F}_{p^m}^*\setminus \{\pm\alpha^{(p^k+1)/2}\}\,|\,\text{both
(\ref{EqM4d-4}) and (\ref{EqM4d-5}) have $2$ solutions}\Big\}.
\end{array}\end{equation} Then (\ref{EqM4bb-N1}) and (\ref{EqM4d-N2}) suggest that $N(\alpha,\beta)=8(p+1)$ if $\beta\in S_1(\alpha)\cap S_2(\alpha)$ and $N(\alpha,\beta)=0$ otherwise.
In what follows, we shall show that if $\alpha$  is a square, then
$S_1(\alpha)\cap S_2(\alpha)=\emptyset$; and if $\alpha$  is a
non-square, then $S_1(\alpha)\subseteq S_2(\alpha)$.

When $\alpha$ is a square, i.e., $\alpha^{(p^m-1)/2}=1$, the
equations in the first row of (\ref{EqM4Equtions}) yield
$$\tau_1^{(p^m+1)/2}=(\theta^{\pm(p^k-1)})^{(p^m+1)/2}=(\theta^{p^m+1})^{\pm(p^k-1)/2}=1,$$
while the equations in the second row of (\ref{EqM4Equtions}) imply
$$\tau_1^{(p^m+1)/2}=(\vartheta^{\pm(p^k+1)})^{(p^m+1)/2}=(\vartheta^{p^m+1})^{\pm(p^k+1)/2}=(-1)^{\pm(p^k+1)/2}=-1.$$
This is a contradiction. Thus, there do not exist $\theta, \vartheta\in
\mathbb{F}_{p^{2m}}^*$ satisfying (\ref{EqM4Equtions}), which is
equivalent to $S_1(\alpha)\cap S_2(\alpha)=\emptyset$.

When $\alpha$ is a non-square, i.e., $\alpha^{(p^m-1)/2}=-1$, one
has $\theta^{p^m+1}=-1$ and $\vartheta^{p^m+1}=1$. Let $\xi$ be a
primitive element of $\mathbb{F}_{p^{2m}}$. Then $\theta$ and
$\vartheta$ can be respectively represented as $
\theta=\xi^{(2i+1)(p^m-1)/2}$ and $\vartheta=\xi^{j(p^m-1)}$, where
$i,j=0,1,\cdots, p^m.$ Assume $\tau_1=\xi^r$, then the equation
$\theta^{p^k-1}=\tau_1$ is equivalent to
$(2i+1)(p^m-1)(p^k-1)/2\equiv r \pmod{p^{2m}-1}.$ For any
$\beta\in S_1(\alpha)$, by the definition of $S_1(\alpha)$, this
linear congruence equation with variable $i$ has $p+1$ solutions.
Thus, $\gcd((p^m-1)(p^k-1)/2,p^{2m}-1)|r.$ Since $\gcd(p^k+1,
p^{m}+1)=2$ and $\gcd((p^k-1)/2, p^{m}+1)=(p+1)$, one has that
$$\gcd((p^m-1)(p^k+1), p^{2m}-1)|\gcd((p^m-1)(p^k-1)/2,
p^{2m}-1)|r.$$ This implies the congruence equation
$j(p^m-1)(p^k+1)\equiv r \pmod{p^{2m}-1}$ with variable $j$
has solutions. Thus, the equations $\vartheta^{p^m+1}=1$ and
$\vartheta^{p^k+1}=\tau_1, \tau^{-1}$ have 4 solutions. This implies
that $\beta$ is also contained in $S_2(\alpha)$. Then $S_1(\alpha)$
is a subset of $S_2(\alpha)$.

Therefore, for  $\alpha\in \mathbb{F}_{p^m}^*$ and $\beta\in
\mathbb{F}_{p^m}^*\setminus \{\pm\alpha^{(p^k+1)/2}\}$, we have
\begin{equation}\label{EqM4-NCaseIIb}
  N(\alpha,\beta)=\left\{
                    \begin{array}{ll}
                      8(p+1), & \hbox{if $\alpha$ is a non-square and $\beta\in S_1(\alpha)$,} \\
                      0, & \hbox{otherwise.}
                    \end{array}
                  \right.
\end{equation}
Combining (\ref{EqM4-N}), (\ref{EqM4-N12CaseI}),
(\ref{EqM4-NCaseII}) and (\ref{EqM4-NCaseIIb}) gives
$$
\mathcal{N}_4=1+((p+1)+4(p+1)|S_1(\alpha)|)(p^m-1).
$$ Thus, to determine the value of $\mathcal{N}_4$, we only need to calculate the cardinality of the set $S_1(\alpha)$. Given $\alpha\in \mathbb{F}_{p^m}^*$,  by \cite[Lemma 6.24]{FiniteFields1997},
the equation $x^2+y^2=\alpha$ has $p^m+1$ solutions in
$\mathbb{F}_{p^m}$. Among all these solutions, by (\ref{EqM4bb-N1}),
if $\beta=\alpha^{(p^k+1)/2}$, there are exactly $p+1$ solutions
satisfying $x^{p^k+1}+y^{p^k+1}=\beta$, and for any $\beta\in
S_1(\alpha)$, there are exactly $2(p+1)$ solutions satisfying
$x^{p^k+1}+y^{p^k+1}=\beta$. Thus, $
(p+1)+2(p+1)|S_1(\alpha)|=p^m+1, $ which implies
$|S_1(\alpha)|=(p^m-p)/2(p+1)$.

Therefore, for $p\equiv 3 \pmod{4}$
$$
\mathcal{N}_4=1+((p+1)+2(p^m-p))(p^m-1)=2p^{2m}-p^{m+1}-p^m+p.
$$
The analysis on the value of $\mathcal{N}_4$ for $p\equiv 1 \pmod{4}$ is similar in spirit to that for $p\equiv 3 \pmod{4}$
and is thus omitted.
The proof is completed. \hfill$\blacksquare$

\end{document}